\documentclass[aps, prd, amsmath, amssymb, amsfonts, floats, %
floatfix, superscriptaddress, nofootinbib, twocolumn, showpacs]%
{revtex4-1}

\allowdisplaybreaks[2]
\usepackage{mathrsfs}
\DeclareMathAlphabet{\mathscrbf}{OMS}{mdugm}{b}{n}
\usepackage{amsmath}
\usepackage{amsbsy}
\usepackage{graphicx}
\usepackage[usenames, dvipsnames]{color}
\usepackage[colorlinks, pdfborder={0 0 0}, plainpages=false]{hyperref}
\usepackage{breakurl}
\usepackage{amsmath, amssymb, amsfonts}
\usepackage{xspace} % Sensible space treatment at end of simple macros
\usepackage{dcolumn}% Align table columns on decimal point
\usepackage{bm}% bold math
\usepackage{multirow} % Multiple rows in tables
\usepackage{times}
\usepackage{mathptmx}

\def\laq{\raise 0.4ex\hbox{$<$}\kern -0.8em\lower 0.62ex\hbox{$\sim$}}
\def\gaq{\raise 0.4ex\hbox{$>$}\kern -0.7em\lower 0.62ex\hbox{$\sim$}}

\definecolor{CiteColor}{rgb}{0, 0.5, 0} %
\hypersetup{citecolor=CiteColor} %
\definecolor{RefColor}{rgb}{0.55, 0, 0} %         
\hypersetup{linkcolor=RefColor} %

\usepackage{ulem}
\definecolor {darkgreen}{rgb}{0.2, 0.7, 0.2}

\newcommand{\MIT}{\affiliation{Department of Physics and MIT Kavli
    Institute, 77 Massachusetts Avenue, Cambridge, MA 02139}}
\newcommand{\Darth}{\affiliation{Department of Physics, University of
    Massachusetts Dartmouth, North Dartmouth, MA 02747}}
\newcommand{\Maryland}{\affiliation{Maryland Center for Fundamental
    Physics \& Joint Space-Science Institute,\\ Department of Physics,
    University of Maryland, College Park, MD 20742, USA}}

\begin{document}

\title{Small mass plunging into a Kerr black hole: 
Anatomy of the inspiral-merger-ringdown waveforms}

\author{Andrea Taracchini} \Maryland %
\author{Alessandra Buonanno} \Maryland %
\author{Gaurav Khanna} \Darth
\author{Scott A.\ Hughes} \MIT%

\begin{abstract}
  We numerically solve the Teukolsky equation in the time domain to
  obtain the gravitational-wave emission of a small mass inspiraling
  and plunging into the equatorial plane of a Kerr black hole. We
  account for the dissipation of orbital energy using the Teukolsky
  frequency-domain gravitational-wave fluxes for circular, equatorial
  orbits, down to the light-ring. We consider Kerr spins $-0.99 \leq q
  \leq 0.99$, and compute the inspiral-merger-ringdown $(2,2)$,
  $(2,1)$, $(3,3)$, $(3,2)$, $(4,4)$, and $(5,5)$ modes. We study the
  large-spin regime, and find a great simplicity in the merger
  waveforms, thanks to the extremely circular character of the
  plunging orbits. We also quantitatively examine the mixing of
  quasinormal modes during the ringdown, which induces complicated
  amplitude and frequency modulations in the waveforms. Finally, we
  explain how the study of small mass-ratio black-hole binaries helps
  extending effective-one-body models for comparable-mass, spinning
  black-hole binaries to any mass ratio and spin
  magnitude.
\end{abstract}

\date{\today}

\pacs{04.25.D-, 04.25.dg, 04.25.Nx, 04.30.-w}

\maketitle

\section{Introduction}
\label{sec:intro}
Over the last few years, analytical and numerical studies have revealed interesting features of the dynamics and 
gravitational radiation of extreme mass-ratio black-hole binaries, especially during ringdown and 
when the spin of the central black hole is close to maximal, and the orbits approach the horizon. 
References~\cite{Castro:2010fd,Hartman:2008pb,Porfyriadis:2014fja,Hadar:2014dpa} pointed out the possibility of describing analytically various processes 
of the dynamics and radiation in the near-horizon region of a nearly extremal black hole by exploiting an infinite-dimensional conformal 
symmetry that the Kerr metric satisfies in this particular limit. Applying the WKB method to the 
Teukolsky equation in the eikonal approximation, Ref.~\cite{Yang:2012he} found a geometric interpretation 
of the black-hole quasinormal modes (QNMs) through spherical light-ring orbits, extending to generic orbits 
what was previously derived for equatorial~\cite{FerrariMashhoon1984,Mashhoon1985} and polar orbits~\cite{Dolan:2010wr}. 
Moreover, an interesting bifurcation leading to a splitting of zero and non-zero damped QNMs was found as one approaches 
nearly-extremal spins~\cite{Yang:2012pj,Yang:2013uba}. Quite interestingly, Refs.~\cite{Mino:2008at, Zimmerman:2011dx} found that damped modes 
different from the usual QNMs are present in the gravitational-radiation spectrum close to the black-hole horizon. It 
remains an open question whether those damped modes are excited as a test body plunges into the central black hole.

Furthermore, gravitational waveforms emitted during the inspiral, plunge and merger stages of a test body orbiting a Kerr black 
hole have been exploited to grasp unique, physical information on the merger phase and they have been employed to 
extend analytical models, notably the effective-one-body (EOB) model~\cite{Buonanno:1998gg,Buonanno:2000ef}, from the comparable-mass to the test-particle limit 
case~\cite{Nagar:2006xv,Damour:2007xr,Bernuzzi:2010ty,Bernuzzi:2010xj,Bernuzzi:2011aj,Barausse:2011kb,
Bernuzzi:2012ku,Taracchini:2012ig,Damour:2012ky,Taracchini:2013wfa}. Solving the time-domain Regge-Wheeler or Teukolsky 
equations is significantly less expensive than evolving a black-hole binary in full numerical relativity. The possibility of 
using the test-particle limit to infer crucial information about the merger waveform of bodies of comparable masses 
follows from the universality of the merger process throughout the binary parameter space. 

In Ref.~\cite{Barausse:2011kb}, some of us investigated the inspiral-merger-ringdown waveforms produced by the time-domain 
Teukolsky equation where the source term is evaluated along the quasicircular plunging trajectory of a nonspinning 
test particle inspiraling in the equatorial plane. The trajectory was computed by solving Hamilton's equations in the Kerr spacetime, augmented by a suitable radiation-reaction force, notably the one constructed from the factorized energy flux 
of the EOB formalism~\cite{Damour:2008gu,Pan:2010hz}. The Teukolsky waveforms were then used to improve spinning EOB waveforms during 
the transition from plunge-merger to ringdown. However, the study of Ref.~\cite{Barausse:2011kb} was limited to moderate spins of 
the Kerr black hole, i.e., $a/M \lesssim 0.8$. Here, we build on Ref.~\cite{Barausse:2011kb}, and extend the analysis in a few directions. First, the analytical energy flux based on spinning, factorized multipolar waveforms~\cite{Damour:2008gu,Pan:2010hz} can differ from the Teukolsky flux; for instance, even for a moderate spin value of 0.7, the modeling error at the innermost stable circular orbit is as large as 10\%. This error comes from a combination of insufficient knowledge of high-order post-Netwonian (PN) terms, and from the truncation at modes with $\ell=8$. As the spin increases, the motion becomes more relativistic and a growing number of modes are excited. Therefore, to overcome this problem, in the equations of motion for the orbital dynamics of the plunging particle, we employ the energy flux computed by a highly-accurate frequency-domain Teukolsky code~\cite{Hughes:1999bq,Drasco:2005kz}. Second, we consider spins in the range $-0.99 \leq a/M \leq 0.99$, but investigate in greater detail 
spins close to extremal, for prograde and retrograde orbits. In fact, those almost-extremal cases display peculiar features 
in the dynamics and waveforms. When the spin is close to 1, the merger waveforms are particularly simple, with a remarkably flat amplitude, as a consequence of the circular nature of the plunge. When the spin is close to $-1$, instead, the phenomenon of QNM mixing dominates the ringdown waveforms. Third, we use those findings to 
suggest a new procedure for modeling the transition from merger to ringdown in the EOB waveforms for spins larger than 
0.8 and mass ratios smaller than $\sim 1/100$. Preliminary results of this paper were employed in Ref.~\cite{Taracchini:2013rva} 
to build a spinning EOB model that is valid for any mass ratio and spin magnitude.
 
This paper is organized as follows. In Sec.~\ref{sec:orbdyn} we describe how we build the orbital dynamics to compute the 
quasicircular plunging trajectory that is used in the source term of the Teukolsky equation. In 
Sec.~\ref{sec:TeukolskyCode} we review the time-domain Teukolsky code which computes the waveforms. In Sec.~\ref{sec:Flat} we describe interesting features 
characterizing the dynamics and the merger waveforms for spins close to extremal. In Sec.~\ref{sec:ModeMix} we carry 
out a detailed study to understand and model the mixing of QNMs for the dominant 
$(2,2)$, $(2,1)$, $(3,3)$, $(3,2)$, $(4,4)$, and $(5,5)$ waveforms. In Sec.~\ref{sec:EOBmodeling} we explain how the information obtained from the Teukolsky waveforms has been used to design a new way 
of generating the EOB merger-ringdown waveform for spins larger than 0.8 and mass ratios smaller than $\sim 1/100$. In 
Sec.~\ref{sec:CompareWaveforms} we compare spinning EOB waveforms developed in the comparable-mass regime~\cite{Taracchini:2013rva} to the Teukolsky waveforms. Sec.~\ref{sec:conclusions} summarizes our main conclusions and discusses future directions. Appendix~\ref{app:InputValues} provides numerical information about the Teukolsky merger waveforms that can be incorporated in generic spinning EOB models.

Henceforth, we use geometric units with $G=c=1$.

\section{Orbital dynamics to generate inspiral-merger-ringdown Teukolsky waveforms}
\label{sec:orbdyn}
\begin{figure*}[!ht]
  \centerline{
    \includegraphics*[scale=0.375]{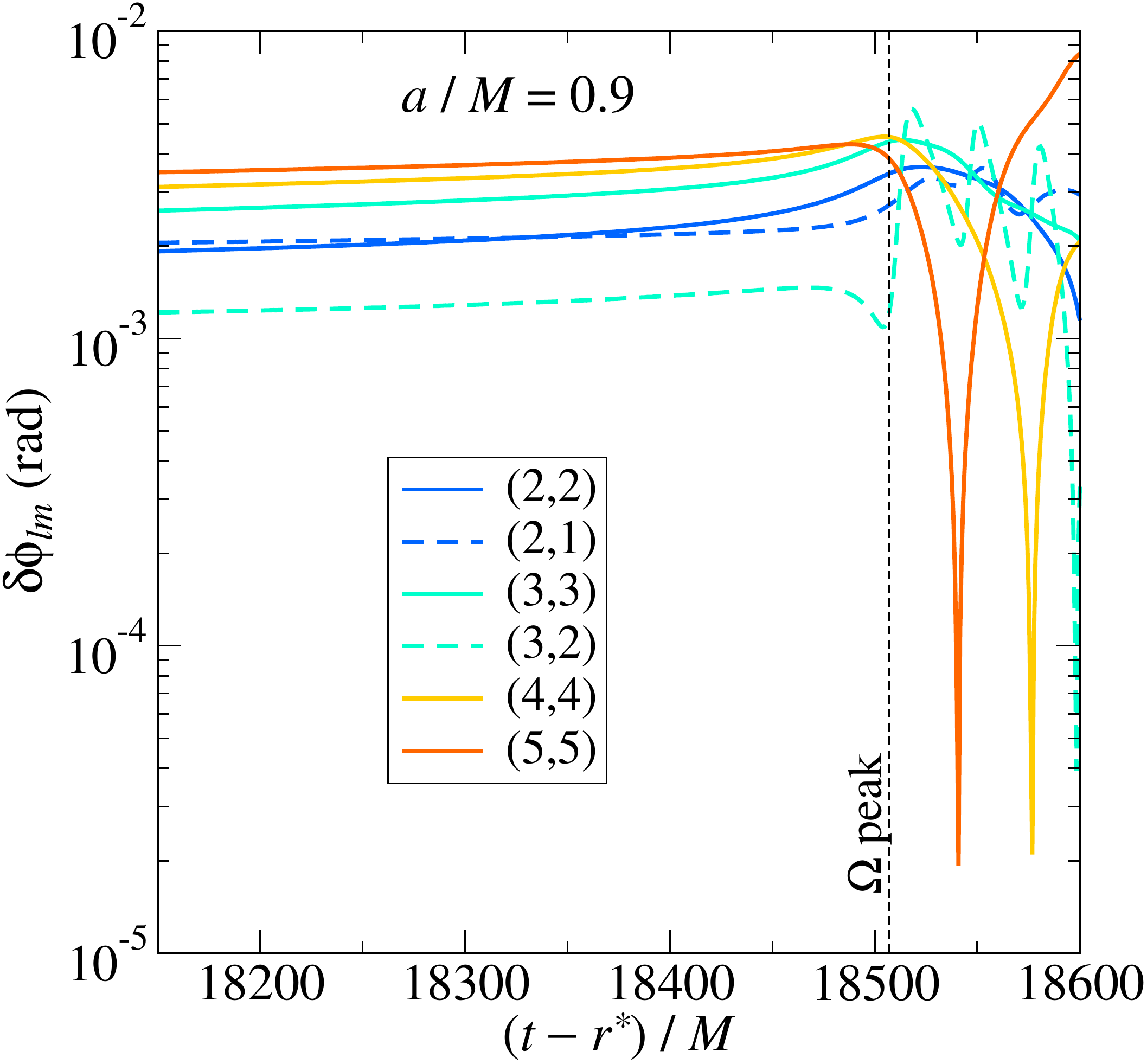}
    \hspace*{2em}
    \includegraphics*[scale=0.375]{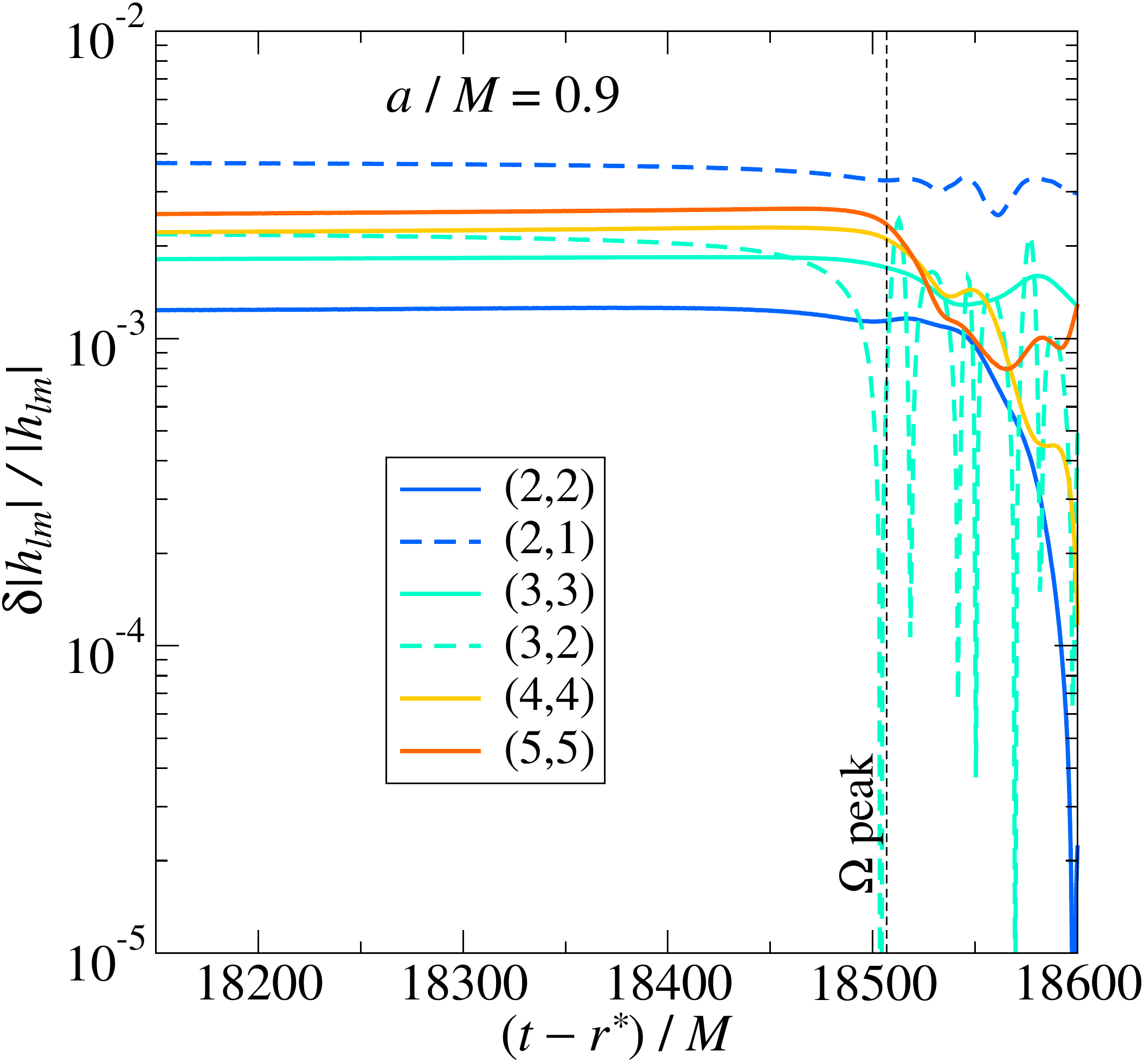}} 
 \caption{\label{fig:NumErr09} Numerical discretization errors in the phase (left panel) and amplitude (right panel) of the Teukolsky waveforms for the $(2,2)$, $(2,1)$, $(3,3)$, $(3,2)$, $(4,4)$, and $(5,5)$ modes. The plots are for spin $q=0.9$. A vertical line marks the position of the peak of the orbital frequency, at time $t_{\rm peak}^{\Omega}$, which occurs close to merger.}
\end{figure*}
In this section we review how the trajectory entering the source term
of the Teukolsky equation is computed. We restrict our attention to
systems where the smaller black hole (BH) is nonspinning, and the orbits are confined to 
the equatorial plane of the larger, spinning BH. Let $\mu$ be the mass of
the smaller object, and let $M$ and $J\equiv a M\equiv q M^{2}$ (with~\footnote{Positive (negative) values of $q$ indicate that the spin of the Kerr BH is aligned (anti-aligned) with the inspiral orbital angular momentum, i.e., the motion is prograde (retrograde) during the inspiral. At the end of the plunge, because of frame dragging, the trajectory always becomes prograde.}
$-1 \leq q \leq 1$) be the mass and spin of the larger one. In this
paper we consider systems with $\mu/M=10^{-3}$. In the spirit of the EOB 
formalism, and as in Ref.~\cite{Barausse:2011kb}, we model the orbital dynamics using the
Hamiltonian of a nonspinning test particle of mass $\mu$ in the Kerr
spacetime
\begin{equation}
H= \beta^{i}p_{i}+\alpha\sqrt{\mu^{2}+\gamma^{ij}p_{i}p_{j}}\,,
\end{equation} 
where $\alpha\equiv (-g^{tt})^{-1/2}$, $\beta^{i}\equiv g^{it}/g^{tt}$
and $\gamma^{ij}\equiv g^{ij}-g^{it}g^{jt}/g^{tt}$, $i,j$ are spatial
indices, $t$ is the time index, $g_{\mu \nu}$ is the Kerr metric in
Boyer-Lindquist coordinates, and the $p_{i}$'s are the conjugate
momenta to the spatial coordinates. We numerically solve Hamilton's
equations for $H$ subject to a radiation-reaction force
${\mathscrbf{F}}$ which describes the dissipation of energy into gravitational waves (GWs); 
the radiation-reaction force is proportional to the sum of the GW energy flux at infinity, $F^{\infty}$,
and through the horizon~\footnote{The GW energy flux falling into the
  horizon is also referred to as ``ingoing flux'', ``absorption
  flux'', or ``horizon flux''.}, $F^{\rm H}$. It reads~\cite{Buonanno:2005xu}
\begin{equation}
\mathscrbf{F}=\frac{F}{\Omega |\bm{r}\times
    \bm{p}|}\bm{p}\,,
\end{equation}
where $F\equiv F^{\infty}+F^{\rm H}$, $\bm{r}$ is the separation vector, and $\Omega \equiv \bm{\hat{J}} \cdot(\bm{r}\times\dot{\bm{r}})/r^2$ is the orbital frequency, where $\bm{\hat{J}}$ is the unit vector along the spin of the Kerr BH. We 
indicate with an over-dot the derivative with respect to time $t$.

Some of us, in Ref.~\cite{Barausse:2011kb}, employed the outgoing
factorized energy flux of Ref.~\cite{Pan:2010hz} for the term
$F^{\infty}$, while setting $F^{\rm H}=0$; that choice was motivated
partly by the focus on understanding the effect of the model flux, and
partly by the availability of numerical Teukolsky energy fluxes only
down to the innermost stable circular orbit (ISCO). Here, instead, we
are mainly interested in the characterization of the Teukolsky
waveforms, and we want to remove any modeling error from the orbital
motion. Similarly to what is done in Ref.~\cite{Han:2011qz}, we source
our equations of motion with GW energy fluxes computed in perturbation
theory; in particular, we use the Teukolsky fluxes of
Ref.~\cite{Taracchini:2013wfa}, where we numerically solved the
Teukolsky equation in frequency
domain~\cite{Hughes:1999bq,Drasco:2005kz} for circular, equatorial
orbits all the way down to a radial separation of $r_{\rm min} =
r_{\rm LR} + 0.01M$, where $r_{\rm LR}/M\equiv
2+2\cos{\left[\frac{2}{3}\arccos{\left(-q\right)}\right]}$ is the
position of the photon orbit, or light-ring (LR)~\cite{Bardeen:1972fi}.  The GW fluxes
were computed for spins from $q=-0.9$ up to +0.9 in steps of 0.1, and
also for $q=\pm 0.95, \pm 0.99$. Those computations assumed circular
orbits, for which a precise relation between radius $r$ and orbital
frequency $\Omega_{\rm circ}$ holds, namely $M\Omega_{\rm circ} =
[(r/M)^{3/2}+q]^{-1}$.

To accurately describe the transition from inspiral to plunge, we adopt here the same strategy used 
in the EOB models of comparable-mass BH binaries~\cite{Damour:2006tr,Boyle:2008ge,Pan:2009wj}. 
First, if we introduce the velocity parameter $v_{\Omega}\equiv (M\Omega)^{1/3}$, 
then the total GW flux for circular orbits can be written as $F =
32\mu^{2}v_{\Omega}^{10}\hat{F}(v_{\Omega})/(5M^{2})$, where 
$\hat{F}(v_{\Omega}) = 1 + \mathcal{O}(v^2_{\Omega})$. Second, we
replace $v_{\Omega}$ in the leading term of $F$ with the non-Keplerian
velocity for a circular orbit defined by $v_{\phi}\equiv \Omega r_{\Omega}$, where $r_\Omega
/M\equiv(M\Omega_{\rm circ})^{-2/3}$ (see also Eq.~(32) in Ref.~\cite{Pan:2009wj}); note
that since we work with nonadiabatic~\footnote{This means that the orbital motion includes not only tangential, 
but also radial velocities.}
orbital evolutions $\Omega \neq \Omega_{\rm circ}$. This replacement moderates the growth of the GW
frequency close to merger~\cite{Damour:2006tr}, and allows a more
accurate modeling of numerical-relativity waveforms in the
comparable-mass regime, also when spins are present~\cite{Pan:2009wj}.

We need to integrate the equations of motion to the event horizon,
$r_+/M \equiv 1 + \sqrt{1 - q^2}$.  Teukolsky fluxes are only
available, however, down to the radius $r_{\rm min} = r_{\rm LR} +
0.01$: Circular orbits do not exist at radii $r < r_{\rm LR}$, and the
growing number of significant multipolar contributions force us to
terminate our flux calculations slightly outside $r_{\rm LR}$ (see
Sec.~IIB of Ref.~\cite{Taracchini:2013wfa} for detailed discussion).
Note that the radial distance between $r_{\rm min}$ and $r_{+}$
decreases from $3M$ when $q=-1$ down to $1M$ when $q=0$, and vanishes
for $q=1$. Therefore, we have to provide a prescription for $\hat{F}$
in the interval $r_{+}<r<r_{\rm min}$.  Since these values of $r$ are
well within the plunge phase, where the conservative part of the
dynamics is known to dominate, we decide to smoothly switch off the GW
flux at $r_{\rm end}=r_{\rm min}$. Let $v_{\rm end}$ be the velocity
of a circular orbit of radius $r_{\rm end}$. Explicitly, if $r<r_{\rm
  end}$ but $v_{\Omega} \leq v_{\rm end}$, then we suppress
$\hat{F}(v_{\Omega})$ by a factor $1/[1 + \exp{[-(r - r_{\rm
        end})/\sigma]}]$; if $r<r_{\rm end}$ and $v_{\Omega} > v_{\rm
  end}$, then we set $\hat{F}(v_{\Omega}) = \hat{F}(v_{\rm end})/[1 +
  \exp{[-(r - r_{\rm end})/\sigma]}]$. We find that, as long as
$\sigma \lesssim 0.01M$, the trajectories are insensitive to the
specific value of $\sigma$. We test the effect of the switch-off point
by changing its position to $r_{\rm end} = r_{\rm LR} + b (r_{\rm
  ISCO} - r_{\rm LR})$, where $r_{\rm ISCO}$ is the position of the
ISCO, and $b=0.25,0.5,0.75$, for spins $q=0.5,0.9$; the difference in
the orbital phase is always negligible, within 0.003 (0.006) rads for
$q=0.5$ (0.9) when $b=0.75$ with respect to the fiducial case $r_{\rm
  end}=r_{\rm min}$ (i.e., $b\approx 0$), since the plunging motion is
indeed geodetic to a good approximation, and is not affected by the
details of the GW fluxes.

As in Ref.~\cite{Barausse:2011kb}, we compute the trajectory from the
equations of motion down to a point slightly outside the horizon (at
$\sim1.05 r_{+}$). Then, to model the locking of the plunging particle 
to the rotating horizon, we smoothly connect the trajectory 
obtained by solving Hamilton's equations to several orbital
cycles at $r=r_{+}$ with frequency equal to that of the horizon
$\Omega_{\rm H} \equiv q/(2 r_{+})$. As shown in
Ref.~\cite{Mino:2008at}, the trajectory asymptotes to $r_{+}$ and
$\Omega_{\rm H}$ exponentially in time.

\section{Numerical solution of the time-domain Teukolsky equation}
\label{sec:TeukolskyCode}

In this section we review the numerical method used to solve the
Teukolsky equation in the time domain. The approach we follow to solve
this linear partial differential equation (PDE) is the same as
presented in our earlier work (see Ref.~\cite{Barausse:2011kb} and
references therein).  The main points of this technique are as
follows: (i) We first rewrite the Teukolsky equation using suitable
coordinates --- the tortoise radius $r^{*}$ and Kerr azimuthal angle
$\varphi$, defined precisely in~\cite{Barausse:2011kb}.  (ii) Taking
advantage of axisymmetry, we separate the dependence on azimuthal
coordinate $\varphi$.  We thus obtain a set of (2+1) dimensional
equations.  (iii) We recast these equations into a first-order,
hyperbolic PDE form.  (iv) Finally, we implement a two-step,
second-order Lax-Wendroff, time-explicit, finite-difference numerical
evolution scheme.  The particle-source term on the right-hand-side of
the Teukolsky equation requires some care for such a numerical
implementation.  All relevant details can be found in our earlier
work~\cite{Barausse:2011kb} and the associated references.

Since Ref.~\cite{Barausse:2011kb} was published, two technical advances have been 
introduced into the solver code aimed at improving results for the present 
paper. First, a compactified hyperboloidal layer has been added to the outer 
portion of the computational domain~\cite{ZengKhanna2011}. This advancement 
allows us to map null infinity onto the computational grid and also 
completely solves the so-called ``outer boundary problem'' (i.e., it 
eliminates unphysical reflections from the artificial boundary of the 
domain). Therefore, differently from Ref.~\cite{Barausse:2011kb}, we are 
now able to extract gravitational waveforms {\em directly} at null infinity, 
completely eliminating the ``extraction error'', as discussed in 
Ref.~\cite{Barausse:2011kb}. Secondly, we have taken advantage of
advances made in parallel computing hardware, and we have developed a very
high-performing OpenCL implementation of the Teukolsky code that takes 
full benefit of GPGPU-acceleration and cluster computing. Details on
this parallel implementation and careful measurements of gains in
overall performance can be found in Ref.~\cite{McKennon:2012iq}.
 
These advances have helped improve the performance and accuracy of the
time-domain Teukolsky code by several orders of magnitude over
previous versions. In particular, Ref.~\cite{McKennon:2012iq}
demonstrated that errors with the improved code are typically at the
level of $0.01\%$, an order of magnitude better than earlier
versions~\cite{Barausse:2011kb}, while performing faster. For long
evolutions, these improvements yield a several thousand-fold
speedup~\cite{ZengKhanna2011}.  Consider the impact of such
improvements on modeling the evolution of a system for $20,000M$, a
typical span for our studies.  With our previous
Cauchy-evolution-based Teukolsky code, we would need to place the
outer boundary at $r \gtrsim 10,000M$ to avoid impact of boundary
effects --- outside the domain of causal influence for the location
and duration of interest.  Using hyperboloidal slicing, the outer
boundary can be placed as close as $50M$~\cite{ZengKhanna2011}.  This
immediately gains two orders of magnitude in performance, while
generating waveforms directly at null infinity as desired.  In
addition, the use of GPGPU compute hardware acceleration typically
yields another order of magnitude gain in performance through
many-core parallelism~\cite{McKennon:2012iq}.

Since we now compute the waveforms exactly at null infinity
(eliminating the extraction error entirely), the only remaining source
of numerical error is the ``discretization error'' introduced by the
finite-difference numerical scheme~\cite{Barausse:2011kb}. It is
relatively straightforward to estimate this discretization error: We
first compute the waveforms at multiple grid resolutions, in
particular we choose $(\textrm{d}r^{*},\textrm{d}\theta) =
(M/80,\pi/128), (M/40,\pi/64)$ and $ (M/20,\pi/32)$. Second, we derive
the Richardson extrapolant using this data. Then, we simply use this
extrapolant as a reference to estimate the discretization error in the
original waveforms computed by our code. In other words, we take the
relative difference between the highest resolution data and the
Richardson extrapolant as a measure of the discretization error. As
done typically in the literature, we decompose the waveforms in
$-2$-spin-weighted spherical harmonic modes, labeled by $(\ell,m)$.
In Fig.~\ref{fig:NumErr09} we depict the discretization errors for the
phase and the amplitude for one particular choice of the spin. These
results should be considered representative of all the other cases
that we present in this work.  Figure~\ref{fig:NumErr09} demonstrates
that the numerical error in our waveform data is at a level of a few
$\times$ $0.1\%$.  As expected, the relative error is generally lower
for the dominant modes such as $h_{22}$ and $h_{33}$, and higher for
the weaker ones. In addition, the error levels stay very uniform
during the long inspiral phase of the binary evolution and only begin
to vary significantly during the plunge.  This happens due to the fact
that the numerical computation shifts from being dominated by the
particle-source term during inspiral, to a nearly source-free
evolution during and after the plunge phase. It should be noted that
the numerical errors can be further reduced by an order of magnitude
as demonstrated in Ref.~\cite{McKennon:2012iq}, through an increase in
grid resolution. However, given the large number and long duration of
the evolutions presented in this work, reducing the numerical error
further was neither very practical nor needed.

\section{Simplicity of inspiral-plunge Teukolsky waveforms for large spins}
\label{sec:Flat}
\begin{figure*}[!ht]
  \begin{center}
    \includegraphics*[width=\textwidth]{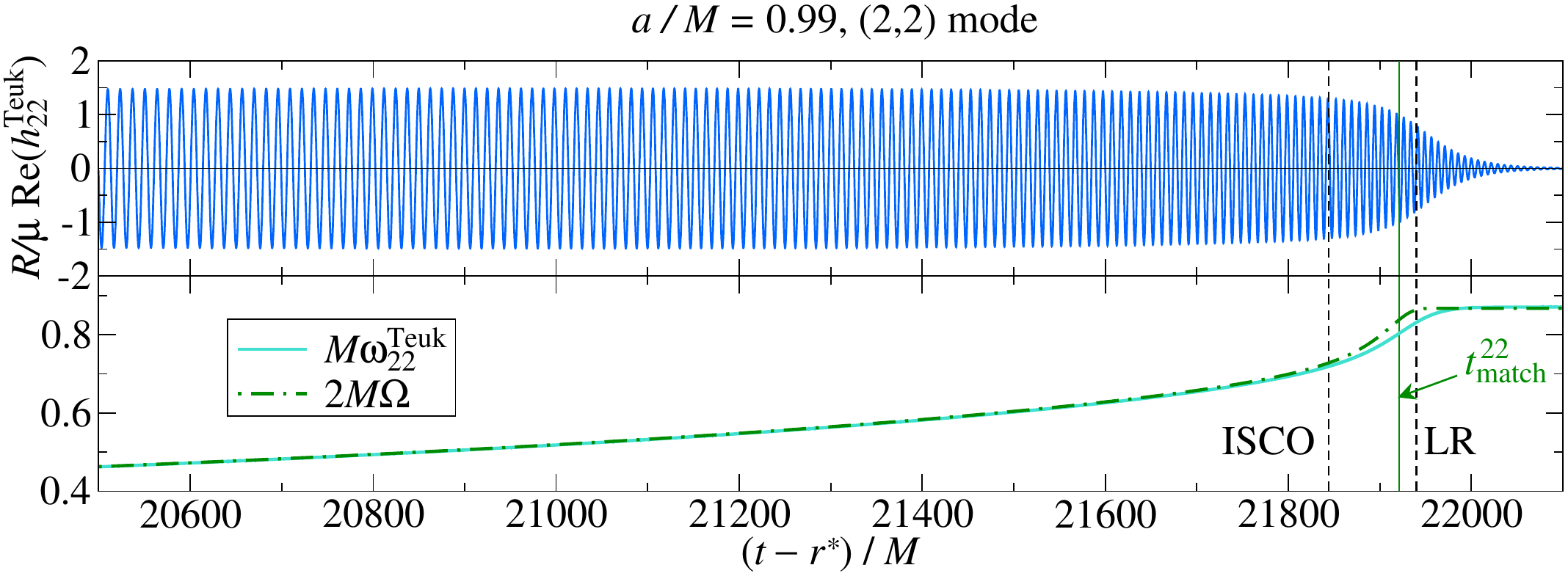}
    \caption{\label{fig:Full22_099} Late inspiral, plunge, merger and ringdown of the Teukolsky $h^{\rm Teuk}_{22}$ waveform (upper panel),
 its GW frequency $\omega^{\rm Teuk}_{22}$ and orbital frequency $\Omega$ of the underlying dynamics (lower panel) for spin $q=0.99$. We note the simplicity of the amplitude during the last 
phase of the evolution. The plot spans a radial range from $r=2.21M$ to the horizon,located at $r=1.14M$; here, $r_{\rm ISCO}=1.45M$ and $r_{\rm LR}=1.17M$. Vertical dashed lines mark the position of the ISCO and the light-ring. A vertical green line marks the position $t_{\rm match}^{22} = t_{\rm peak}^{\Omega} + \Delta t_{\rm peak}^{22}$ of the ringdown matching as prescribed in the EOB model of Ref.~\cite{Taracchini:2013rva} (see discussion in Sec.~\ref{sec:EOBmodeling}). $R$ is the distance to the source.}
  \end{center}
\end{figure*}
In this section we characterize the salient features displayed by the
Teukolsky waveforms during late inspiral and plunge. For spins
  $q=0$, $\pm 0.5$, $\pm 0.7$, $\pm 0.8$, $\pm 0.9$, $\pm 0.95$, and $\pm 0.99$, we compute the Teukolsky $(2,2)$, $(2,1)$,
  $(3,3)$, $(3,2)$, $(4,4)$, and $(5,5)$ modes as explained in
  Secs.~\ref{sec:orbdyn} and \ref{sec:TeukolskyCode}. In the
  test-particle limit, as the spin increases, more and more $(\ell,m)$
  modes become important at merger (with respect to the $(2,2)$ mode); however,
  for the modeling of comparable-mass BH binaries, only the few modes
  above give significant contribution to the energy flux. Eventually, we are 
interested in exploiting the results of this paper in the comparable-mass 
limit, therefore we restrict the discussion to those modes.

In the large-spin regime, a prograde inspiraling particle reaches very
relativistic speeds before getting to the horizon; for instance, when
$q=0.99$, the peak speed (attained at the peak of the orbital
frequency) is around 0.75. At such speeds, the PN expansion 
is inadequate for analytically describing such systems.  However, the
Teukolsky inspiral-merger waveforms turn out to be extremely
simple. Consider, for example, the $(2,2)$ mode emitted when
$q=0.99$, shown in Fig.~\ref{fig:Full22_099}. The prominent feature
that we recognize is the extreme flatness of the amplitude versus
time, across hundreds of $M$, well before the plunge starts at the
ISCO. The GW frequency $\omega^{\rm Teuk}_{22}$, defined as $-\Im{(\dot{h}_{22}^{\rm Teuk}/h_{22}^{\rm Teuk})}$, 
does not display any particular characteristic, and we notice that it is well approximated
by twice the orbital frequency even during ringdown, thanks to the
fact that $2\Omega_{\rm H}$ is very close to the least-damped
quasinormal mode. We find that the flattening of the amplitudes
$|h^{\rm Teuk}_{\ell m}|$ around their respective peaks is more and
more apparent as $q$ approaches 1. This aspect of the numerical
waveforms does not depend on minute details of the flux used to
generate the underlying orbital dynamics.
\begin{figure}[!ht]
  \begin{center}
    \includegraphics*[width=0.42\textwidth]{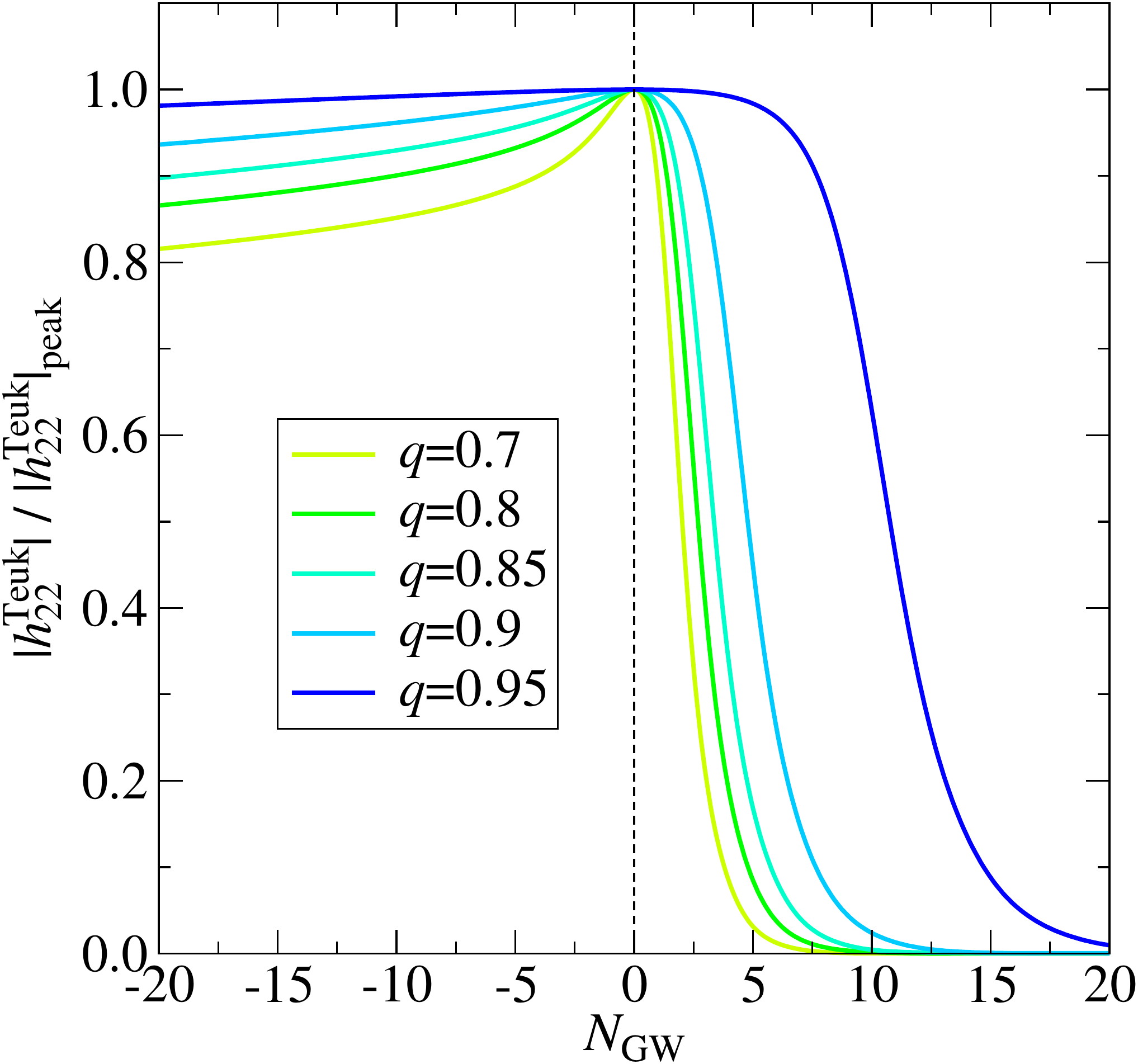}
    \caption{\label{fig:Flat22LargeSpin} Flattening of the peak amplitude of the Teukolsky $(2,2)$ mode as the spin grows towards 1. The curves are normalized by the values of the amplitude at the peak. We align the waveforms in time at $t_{\rm peak}^{22}$, and plot them as functions of the number of GW cycles from the peak.}
  \end{center}
\end{figure}
In Fig.~\ref{fig:Flat22LargeSpin} we show the amplitudes of the Teukolsky $(2,2)$ modes for $q=0.7,0.8,0.85,0.9,0.95$ aligned at $t_{\rm peak}^{22}$, with $t_{\rm peak}^{\ell m}$ being the time when the $(\ell, m)$ mode reaches its maximum amplitude. The almost extremal case $q=0.99$ was not included in Fig.~\ref{fig:Flat22LargeSpin} since its $(2,2)$ amplitude is so flat that it is quite difficult to localize $t_{\rm peak}^{22}$.  In fact, across the $(2,2)$ peak, over a large time interval, its $\partial_{t}|h_{22}^{\rm Teuk}|$ is so small that it is dominated by numerical noise, making it difficult to clearly locate its zero. The curvature $(\partial^2_t |h_{22}^{\rm Teuk}|)_{\rm peak}$ becomes vanishingly small as $q \to 1$; see also Fig.~\ref{fig:IVAmp}. Although we have shown only $(2,2)$ mode waveforms, the same holds true for higher harmonics.

\begin{figure}[!ht]
  \begin{center}
    \includegraphics*[width=0.43\textwidth]{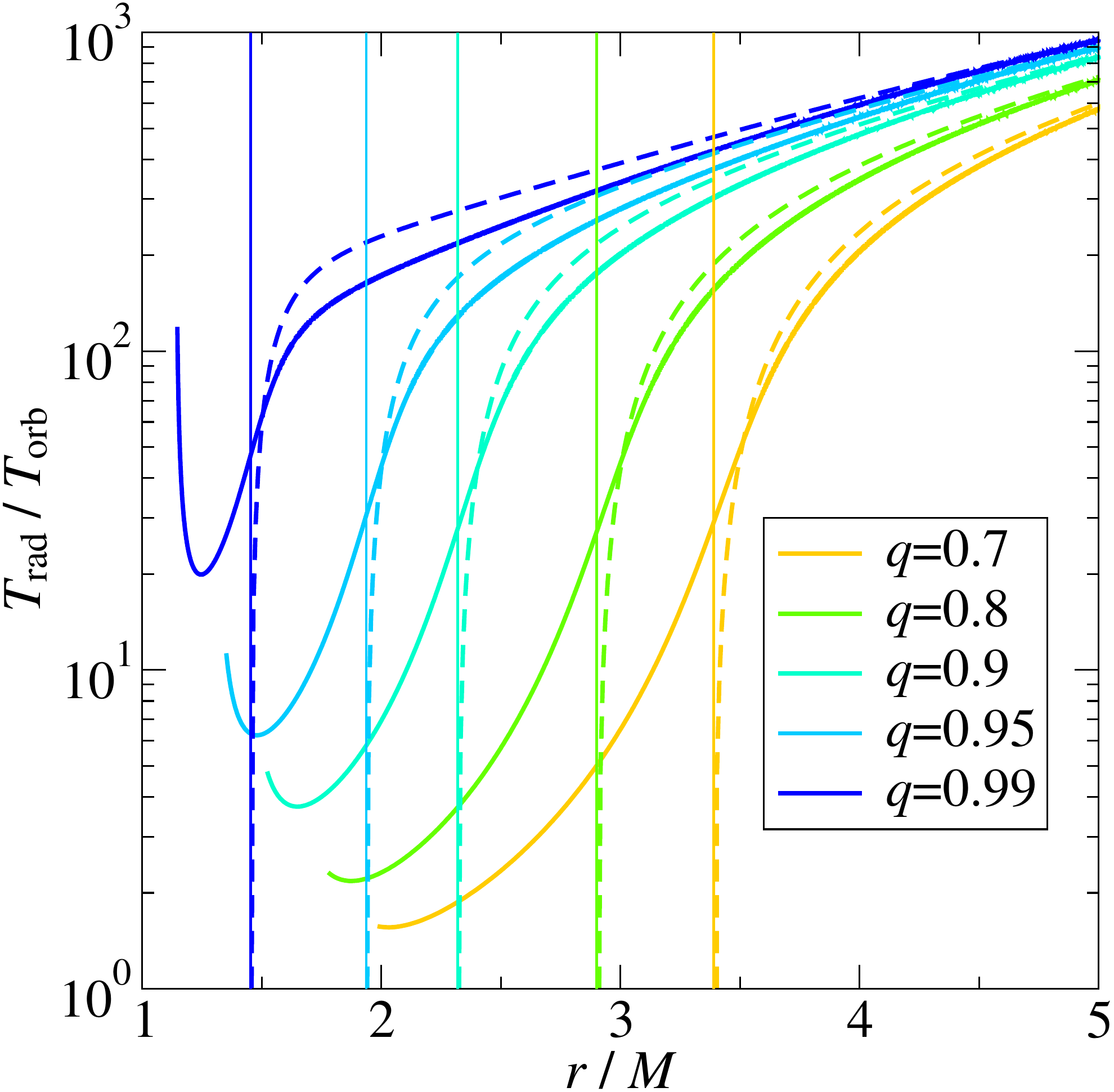}
    \caption{\label{fig:Timescales} Ratio between the radiation-reaction timescale $T_{\rm rad}$ and the orbital period $T_{\rm orb}$ as a function of the radial separation for large positive spins. The curves extend up to the peak of the orbital frequency. The solid lines are computed from the numerical integration of the equations of motion. The dashed lines are the analytical predictions for the quasicircular regime, using only quadrupolar emission (Eq.~(\ref{Trad})). Vertical lines mark the position of the respective ISCOs.}
  \end{center}
\end{figure}

\begin{figure}[!ht]
  \begin{center}
    \includegraphics*[width=0.45\textwidth]{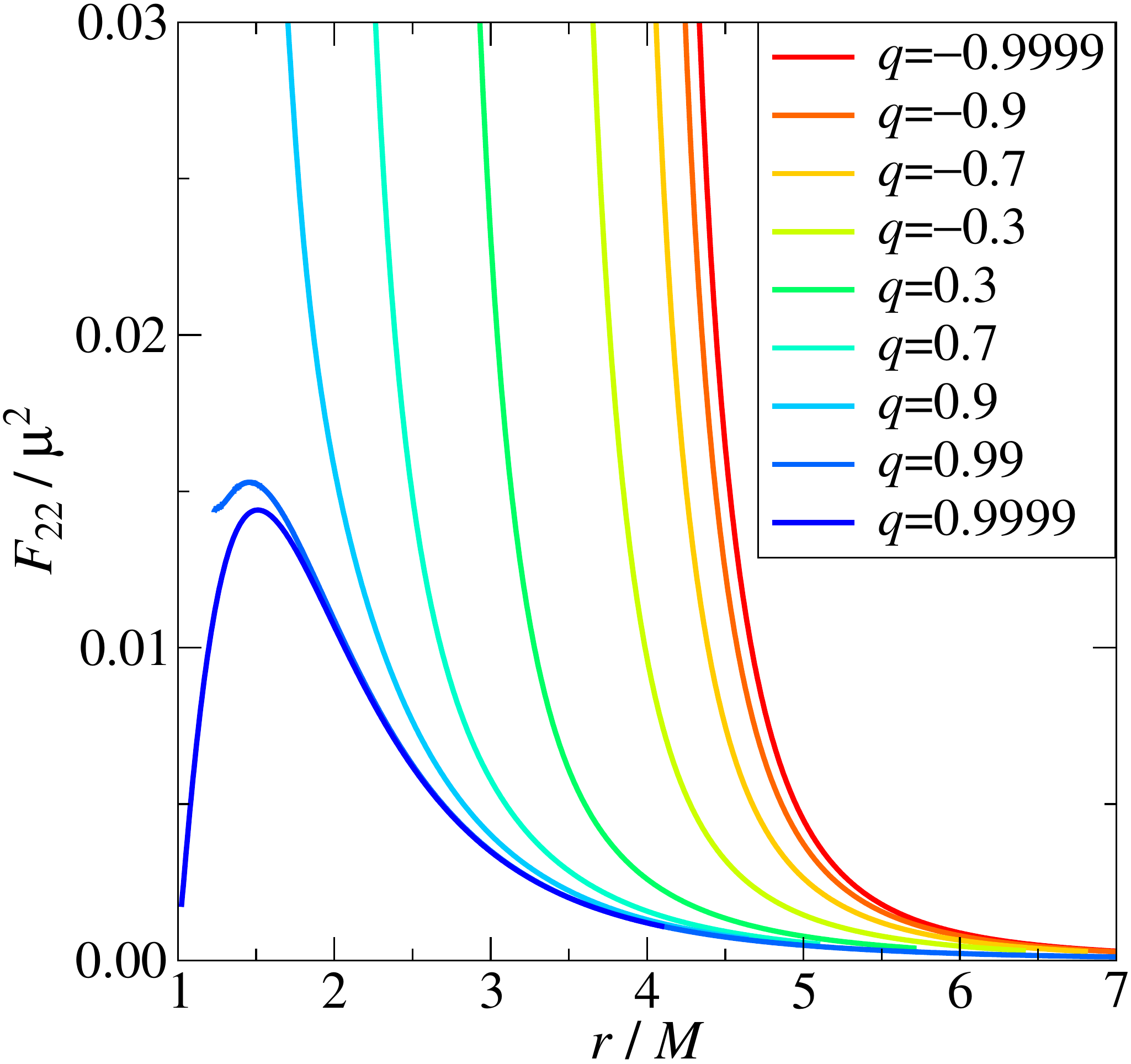}
    \caption{\label{fig:PlungeF22} $(2,2)$ component of the ingoing $+$ outgoing Teukolsky GW flux. The curves extend down to $r_{\rm min} = r_{\rm LR} + 0.01M$.}
  \end{center}
\end{figure}

We can find a physical explanation for why this happens considering the underlying orbital dynamics. As the spin grows larger, the ISCO moves to smaller separations and gets closer to the horizon, so that the plunging phase becomes shorter (in the radial coordinate), and moves to higher frequencies.  This is equivalent to saying that Kerr BHs with larger spins support longer quasicircular inspirals given the same initial frequency. For instance, let us consider spins 0.5 and 0.99. Their dimensionless horizon frequencies are 0.13 and 0.43 respectively. An initial orbital frequency of 0.1 corresponds to radial separations, $4.5M$ and $4.3M$, respectively, 
which are quite close to each other; while for spin 0.5 we are sitting just outside the ISCO ($r_{\rm ISCO}(q=0.5)=4.2M$), for spin 0.99 we are still far from it ($r_{\rm ISCO}(q=0.99)=1.5M$). Furthermore, for very large spins the orbital timescale $T_{\rm orb}$ is much shorter than the radiation-reaction timescale $T_{\rm rad}$. We can estimate these characteristic timescales for different values of $q$ as $T_{\textrm{orb}}=2\pi/\Omega$ and $T_{\textrm{rad}}=-r/\dot{r}$. The orbital frequency grows during the inspiral, reaches a peak value $\Omega_{\rm peak}$ at time $t_{\rm peak}^{\Omega}$, and eventually converges to the horizon frequency $\Omega_{\rm H}$ at late times. One can show that, for all practical purposes, the peak of $\Omega$ occurs at a radius $r_{\rm peak}^{\Omega}$ which nearly coincides with $r_{\rm LR}$, the coincidence being exact for $q=0,1$. In Fig.~\ref{fig:Timescales} we plot the ratio $T_{\rm rad}/T_{\rm orb}$ as a function of the radial separation $r$. The solid lines are computed along nonadiabatic trajectories from the numerical integration of the equations of motion, up to the peak of the orbital frequency $\Omega$. At fixed $r$, the orbital timescale $T_{\rm orb}$ does not vary much with $q$: for example, when $r=4M$, $T_{\rm orb}=53M$ for $q=0.5$, while $T_{\rm orb}=56M$ for $q=0.99$; but the ratio $T_{\rm rad}/T_{\rm orb}$ for spin 0.99 is 55 times larger than for spin 0.5. Hence, the plot demonstrates that there is a clear hierarchy in the radiation-reaction timescales: the larger the spin, the larger $T_{\rm rad}$. As a result, the secular evolution is much slower for large spins, given the same initial separation. This hierarchy can be easily understood using analytical considerations at 
leading order. 
During the quasicircular inspiral we have $T_{\rm orb}\approx 2\pi/\Omega_{\rm circ}$ and the orbital energy $E$ can be approximated by the energy of a circular orbit in Kerr spacetime~\cite{Bardeen:1972fi}
\begin{equation}\label{Ecirc}
\frac{E_{\rm circ}}{\mu} = \frac{1-2M/r + q (M/r)^{3/2}}{\sqrt{1 - 3M/r + 2 q (M/r)^{3/2}}}\,.\end{equation}
Note that $E_{\rm circ}$ diverges at $r=r_{\rm LR}$. Moreover, assuming mainly leading quadrupolar energy loss~\cite{Maggiore2008} and circularity, we get $F=-\dot{E}\approx 32\mu^{2}r^{4}\Omega_{\rm circ}^{6}/5$; thus, we find that
\begin{equation}
T_{\rm rad}=-r\frac{\textrm{d}E/\textrm{d}r}{\textrm{d}E/\textrm{d}t}\approx\frac{\textrm{d}E_{\rm circ}/\textrm{d}r}{\frac{32}{5}\mu^{2}r^{3}\Omega_{\rm circ}^{6}}\,.
\label{Trad}
\end{equation}
In Fig.~\ref{fig:Timescales} we plot the analytical estimate (\ref{Trad}) with dashed lines, and find that it captures the
  numerical result (solid lines) fairly well at large $r$, and, most importantly, can
  account for the hierarchy of the curves due to the presence of
  spin. We can now understand why large-spin waveforms are so
  flat. For large $q$ the radiation-reaction timescale is much larger
  than the orbital timescale, which means that the particle performs
  many orbits while sweeping very slowly through the frequency range up
  to the horizon, so that the secular evolution of the emitted GW
  signal is much slower as compared to systems with smaller $q$. This is consistent with the behavior of the frequency-domain Teukolsky fluxes that we employ in the equations of motion, whose $(2,2)$ component is plotted in Fig.~\ref{fig:PlungeF22} versus radius; at fixed $r$, the dissipation of energy is smaller for larger spins.
\begin{figure}[!ht]
  \begin{center}
    \includegraphics*[width=0.45\textwidth]{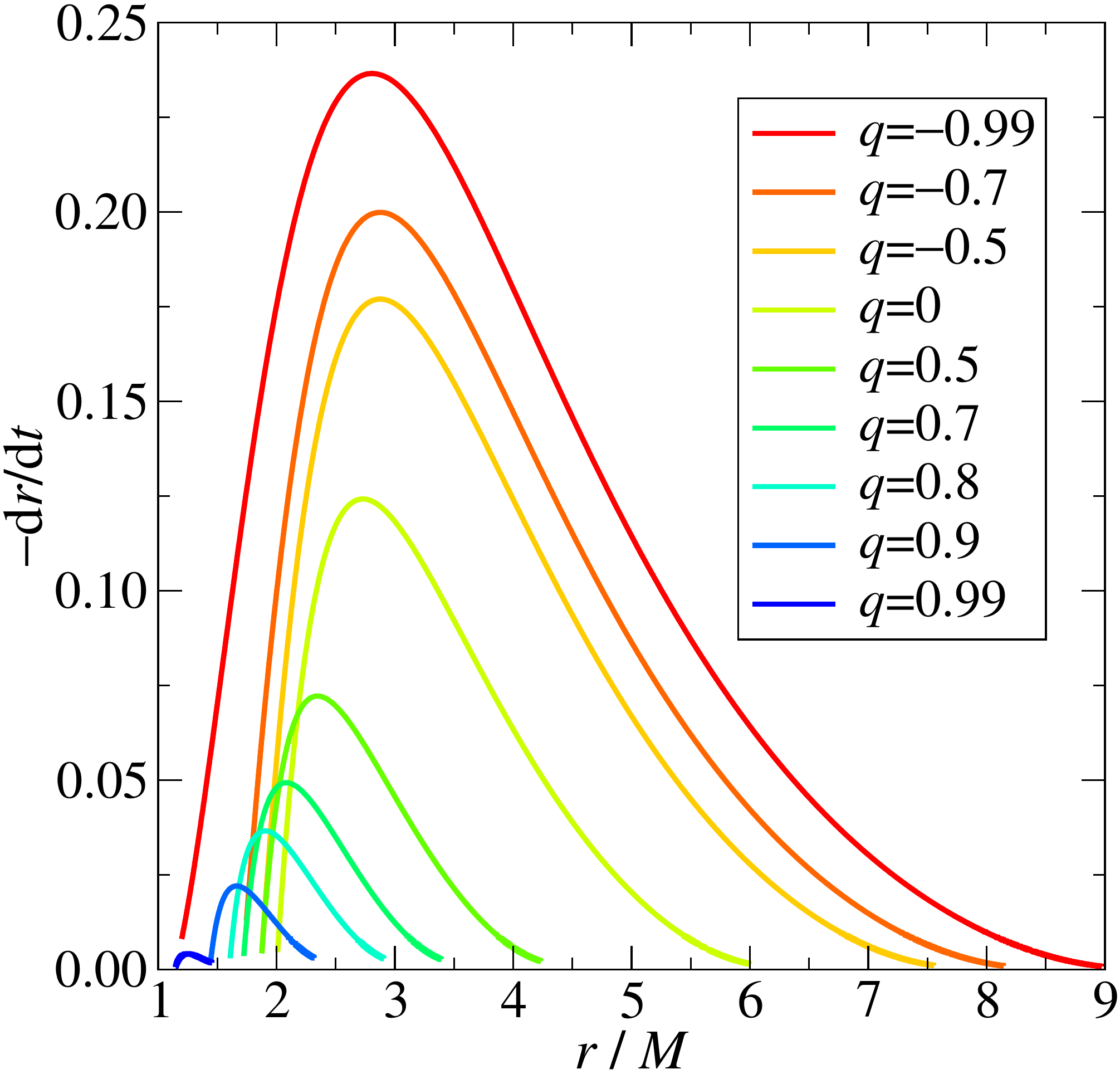}
    \caption{\label{fig:PlungeVel} Radial velocity $\dot{r}$ during the plunge. The curves cover the range $r_{+} < r < r_{\rm ISCO}$.}
  \end{center}
\end{figure}

 Notice that, only for this plot, we include spins as large as $q=\pm0.9999$. Interestingly, as $q\to 1$ the fluxes become small even outside the ISCO and approach vanishingly small values beyond the ISCO, which accounts for the behavior of the ratio $T_{\rm rad}/T_{\rm orb}$ in the late inspiral and plunge. We notice that, starting from $q=0.99$, $F_{22}$ does not display the characteristic divergence at the light-ring as $(E_{\rm circ}/\mu)^{2}\sim (r - r_{\rm LR})^{-1}$, which is well known~\cite{Davis:1972dm,Breuer:1973kt, Chrzanowski:1974nr,BreuerBook}. Instead, $F_{22}$ tends to decrease towards 0, and, remarkably, becomes linear in $(r-r_{+})$ for $q=0.9999$, when 
$r_{\rm LR} \sim r_+$. This is in agreement with analytical work on the gravitational radiation from a particle plunging into a nearly-extremal Kerr BHs in Ref.~\cite{Porfyriadis:2014fja}.

Furthermore, during the plunge (which is governed mostly by conservative effects), the radial velocity $\dot{r}$ reaches maximum values that decrease with $q$, meaning that for large spins even the plunge is not too far from being circular. Figure~\ref{fig:PlungeVel} plots the $r$--dependence of the radial velocity in the region inside the ISCO for several different spin configurations, using orbital evolutions obtained by 
solving Hamilton's equations.  The peak radial velocity differs by more than one order of magnitude between $q=-0.99$ and $q=0.99$.

\section{Quasinormal-mode mixing in ringdown Teukolsky  waveforms and its modeling}
\label{sec:ModeMix}
\begin{figure}[h!]
    \includegraphics[width=\linewidth]{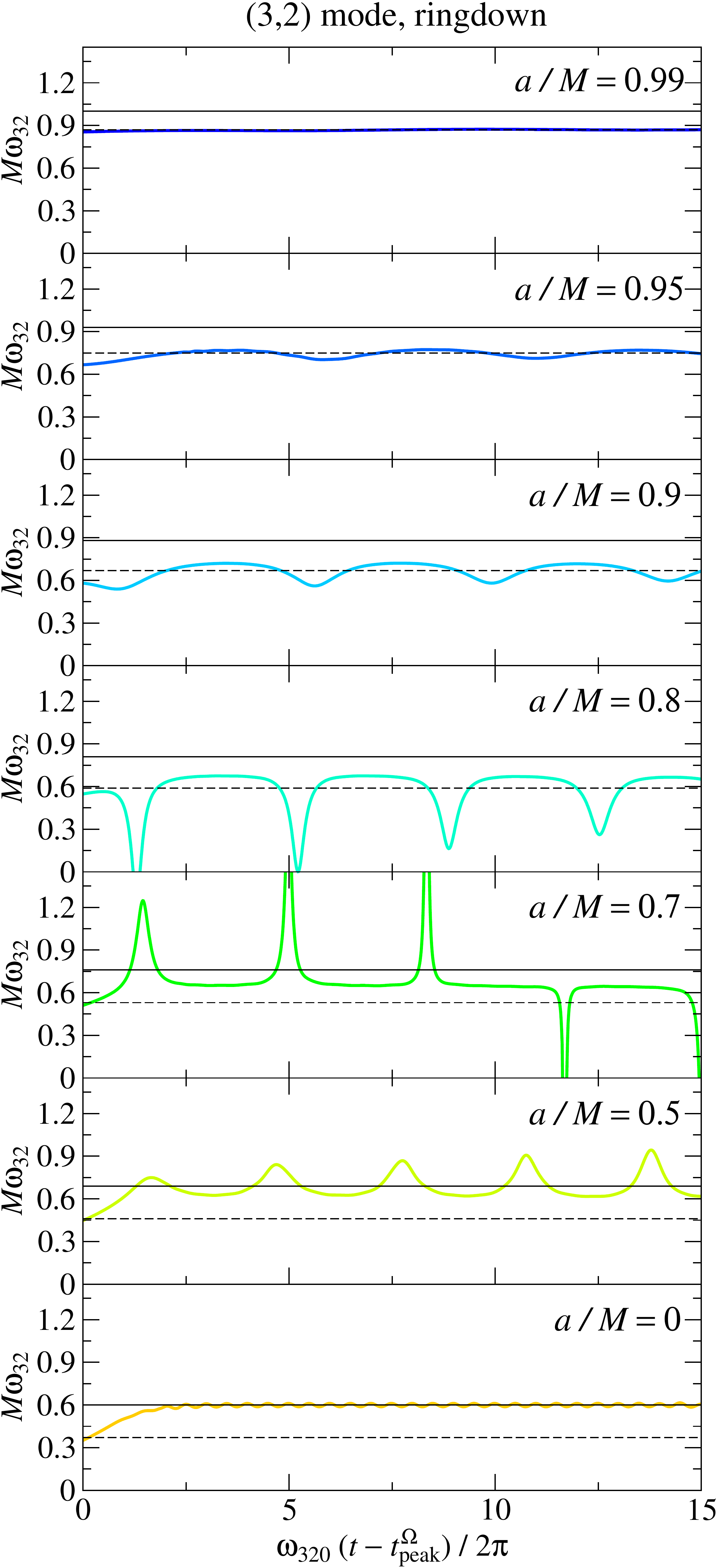}
 \caption{\label{fig:Stack32} GW frequencies of Teukolsky $(3,2)$ ringdown waveforms with positive spin. For the common $x$-axis, we use the time elapsed from the orbital frequency peak in units of $2\pi/\omega_{320}$. Solid horizontal lines indicate the value $M\omega_{320}$, while dashed horizontal lines indicate the value $M\omega_{220}$.}
\end{figure}
The merger of a BH binary (of any mass ratio) eventually leads to the formation of a remnant Kerr BH of mass $M_{f}$ and dimensionless spin $q_{f}$. In this work, since we are dealing with an extreme mass-ratio system, we have $M_{f}=M$ and $q_{f}=q$. In the process of settling down to its final, stable state, the binary emits GWs. Those waves can be modeled as a linear superposition of quasinormal modes (QNMs)~\cite{Berti:2005ys,Berti:2009kk} with complex frequencies $\sigma_{\ell m n}$, which depend only on $M_{f}$ and $q_{f}$, and are labelled by the spheroidal-harmonic indices $(\ell,m)$ and by an overtone index $n=0,1,\cdots$. For future convenience, we define $\omega_{\ell m n}\equiv \Re{(\sigma_{\ell m n})}$ and $\tau_{\ell m n}\equiv -1/\Im{(\sigma_{\ell m n})}$. We adopt the convention that $\omega_{\ell m n}>0$ and $\tau_{\ell m n}>0$ for any choice of the indices $(\ell,m,n)$.

In general, the strain waveform $h$ during the ringdown (RD) contains QNMs with all possible values of $(\ell,m,n)$. Additionally, given a spin $q$ and indices $(\ell,m)$, the angular differential equation which stems from the separation of the Teukolsky equation in spheroidal coordinates admits a pair of solutions characterized by frequencies $\sigma_{\ell\pm m n}$. This implies that, whenever considering a specific component $(\ell,m)$ of $h$, even in principle, we get contributions from both positive- and negative-$m$ modes. As argued by Ref.~\cite{Berti:2005ys}, restricting to only positive-$m$ modes would enforce the assumption of circular polarization of the radiation. Of course, the actual importance of the modes depends on the details of how they are excited by the perturbing source, and by their decay times. 

As already found by numerical investigations of the extreme and small mass-ratio
limits~\cite{Krivan:1997hc,Dorband:2006gg,Nagar:2006xv,Damour:2007xr,Bernuzzi:2010ty,Barausse:2011kb},
the dominant and leading subdominant ringdown Teukolsky modes can
display a rich amplitude and frequency structure that hints at the interference of different
QNMs besides the overtones of the least-damped mode, a phenomenon known as \emph{mode mixing}. On the contrary, in the case of comparable-mass BH binaries, mode mixing seems 
less ubiquitous, and so far it has only been seen in the  
$(3,2)$ mode~\cite{Buonanno:2006ui,Berti:2007fi,Schnittman:2007ij,Baker:2008mj,Pan:2011gk,Kelly:2011bp,Kelly:2012nd}. 
For this reason, in the past, when modeling the ringdown 
of the $(\ell,m)$ mode in the EOB approach, one could simply use the $(\ell,m,n)$ QNMs. However, the lack of mode mixing during ringdown in the 
comparable-mass case is inferred by the analysis of nonspinning, nonprecessing or mildly 
precessing configurations. We do not know yet whether this conclusion will hold when strongly precessing 
systems with mass ratios $\gaq\,1/10$ will be considered.

\begin{figure*}
  \centerline{
    \includegraphics*[scale=0.375]{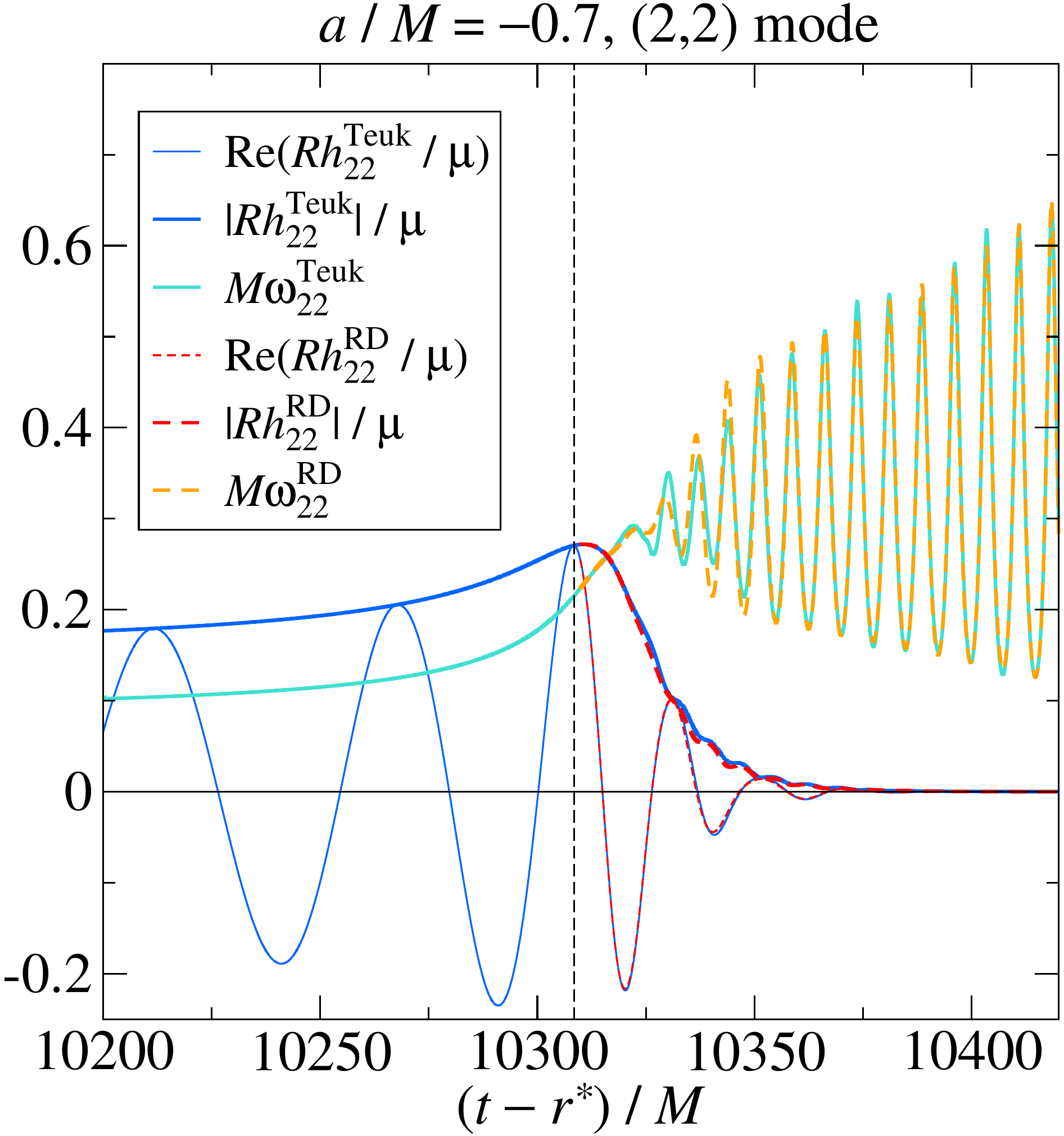}
    \hspace*{4em}
    \includegraphics*[scale=0.375]{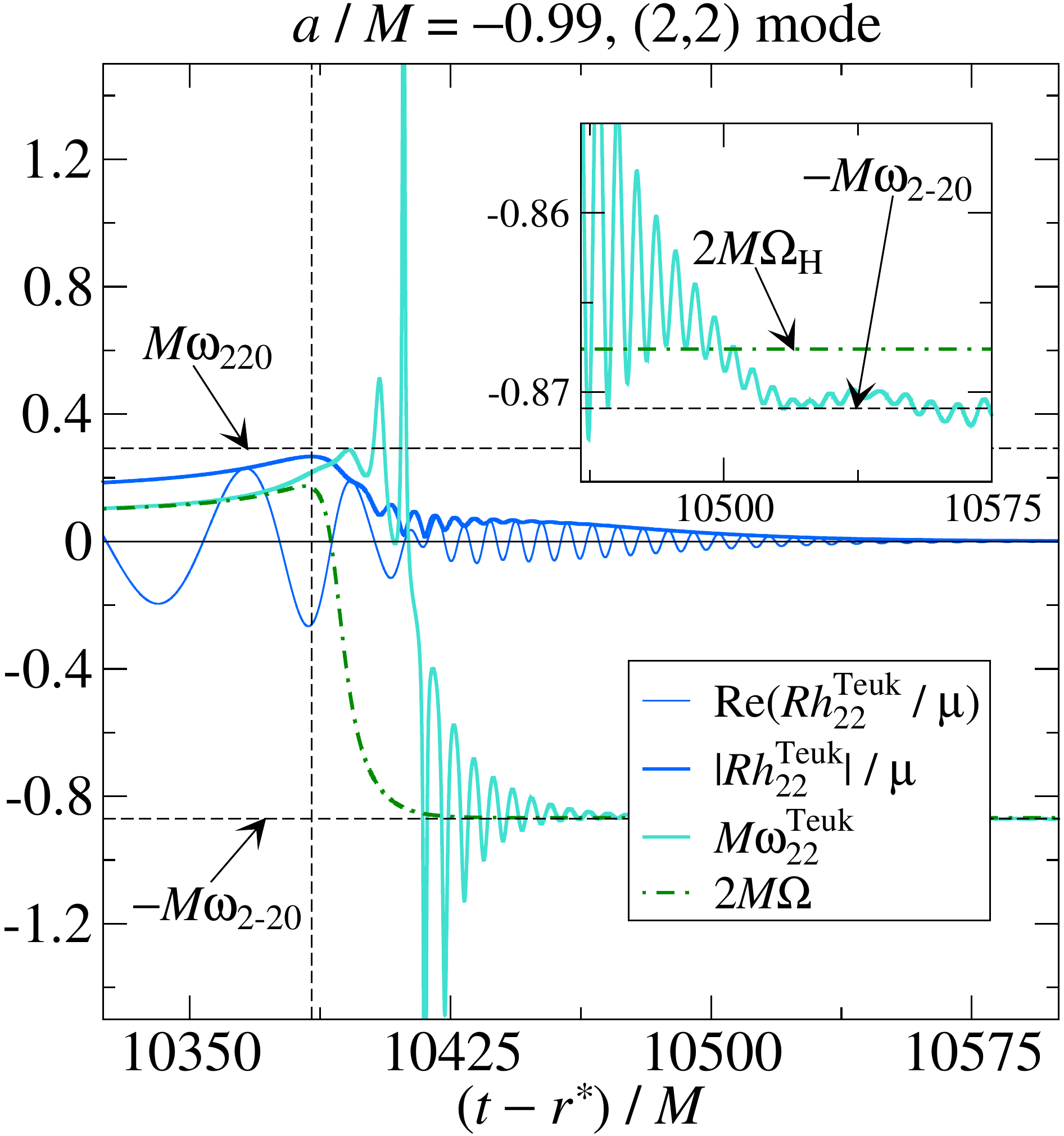}} 
 \caption{\label{fig:a_-07_22} Teukolsky $(2,2)$ mode waveforms for spin $q=-0.7$ (left panel) and $-0.99$ (right panel), displaying mode mixing during the ringdown phase. For $q=-0.7$ we plot a ringdown waveform which contains the mode $(2,-2,0)$ besides the usual $(2,2,n)$ ($n=0,1,\cdots$). The vertical dashed lines mark $t=t_{\rm match}^{22}$. Note that the amplitudes have been rescaled by a factor of 5. $R$ is the distance to the source.}
\end{figure*}

\begin{figure*}
  \centerline{
    \includegraphics*[scale=0.375]{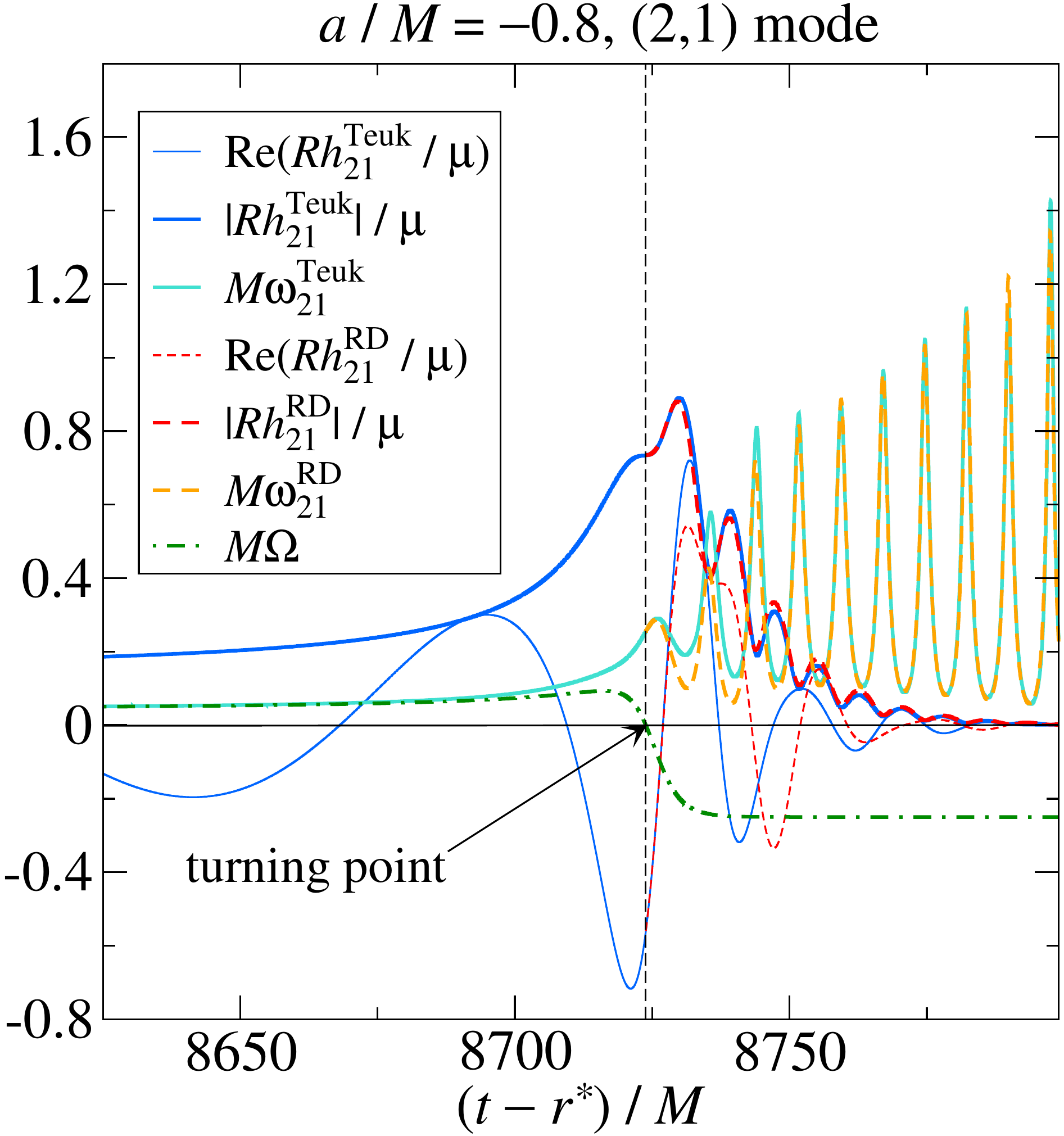}
    \hspace*{4em}
    \includegraphics*[scale=0.375]{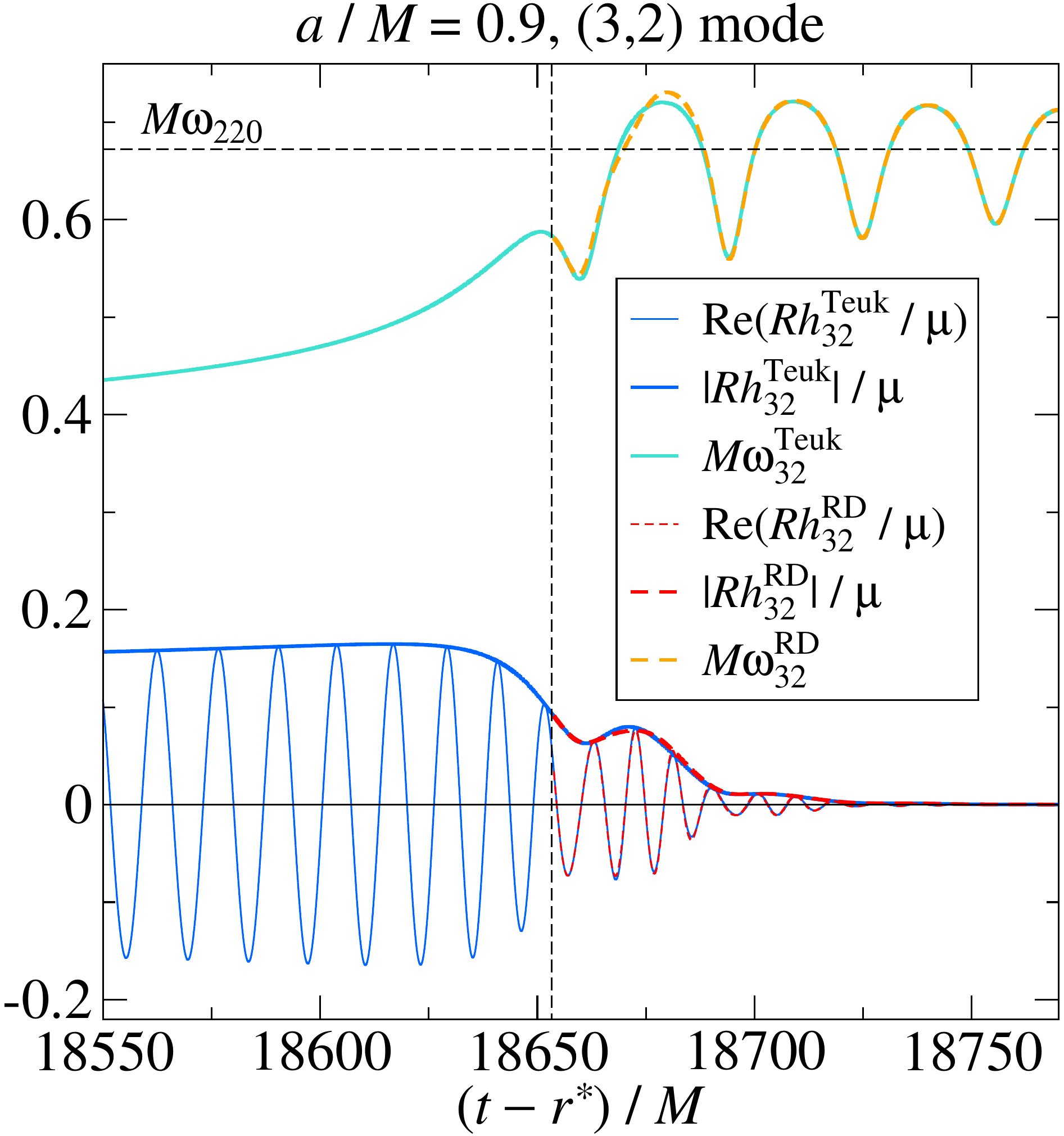}}
 \caption{\label{fig:a_-08_21} Teukolsky $(2,1)$ mode for spin $q=-0.8$ (left panel) and $(3,2)$ mode for spin $q=0.9$ (right panel). For the $q=-0.8$ waveform, the modulations come from the interference of $(2,1,0)$ and $(2,-1,0)$; note how the amplitude peak is affected by the mixing, starting at a time where $\Omega=0$, i.e., the turning point of the azimuthal motion. For the $q=0.9$ waveform, instead, the ringdown contains the modes $(3,2,n)$'s and $(2,2,0)$.The vertical dashed lines mark $t=t_{\rm match}^{\ell m}$. $R$ is the distance to the source.}
\end{figure*}

\begin{figure*}
  \centerline{
    \includegraphics*[scale=0.375]{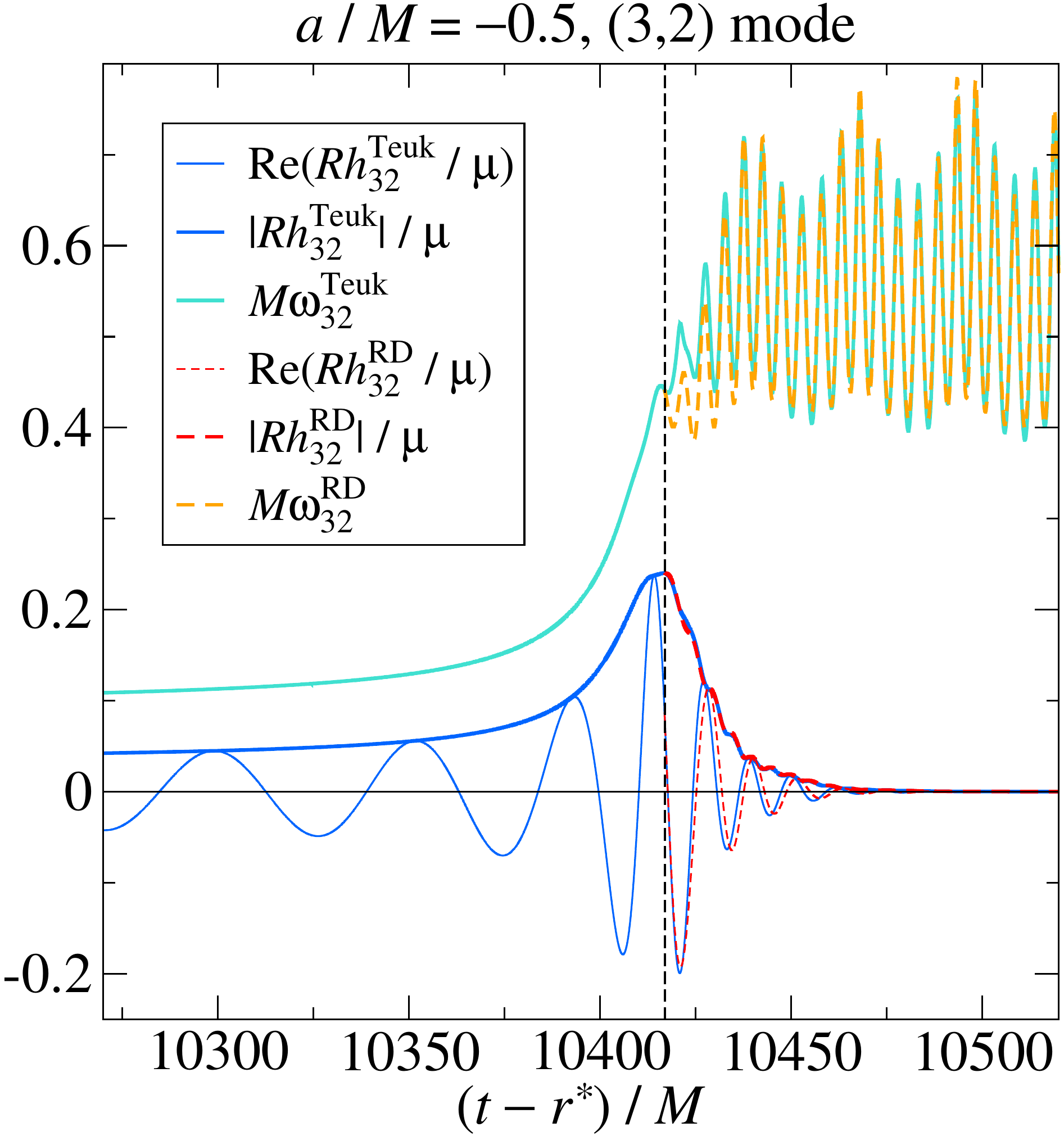}
    \hspace*{4em}
    \includegraphics*[scale=0.375]{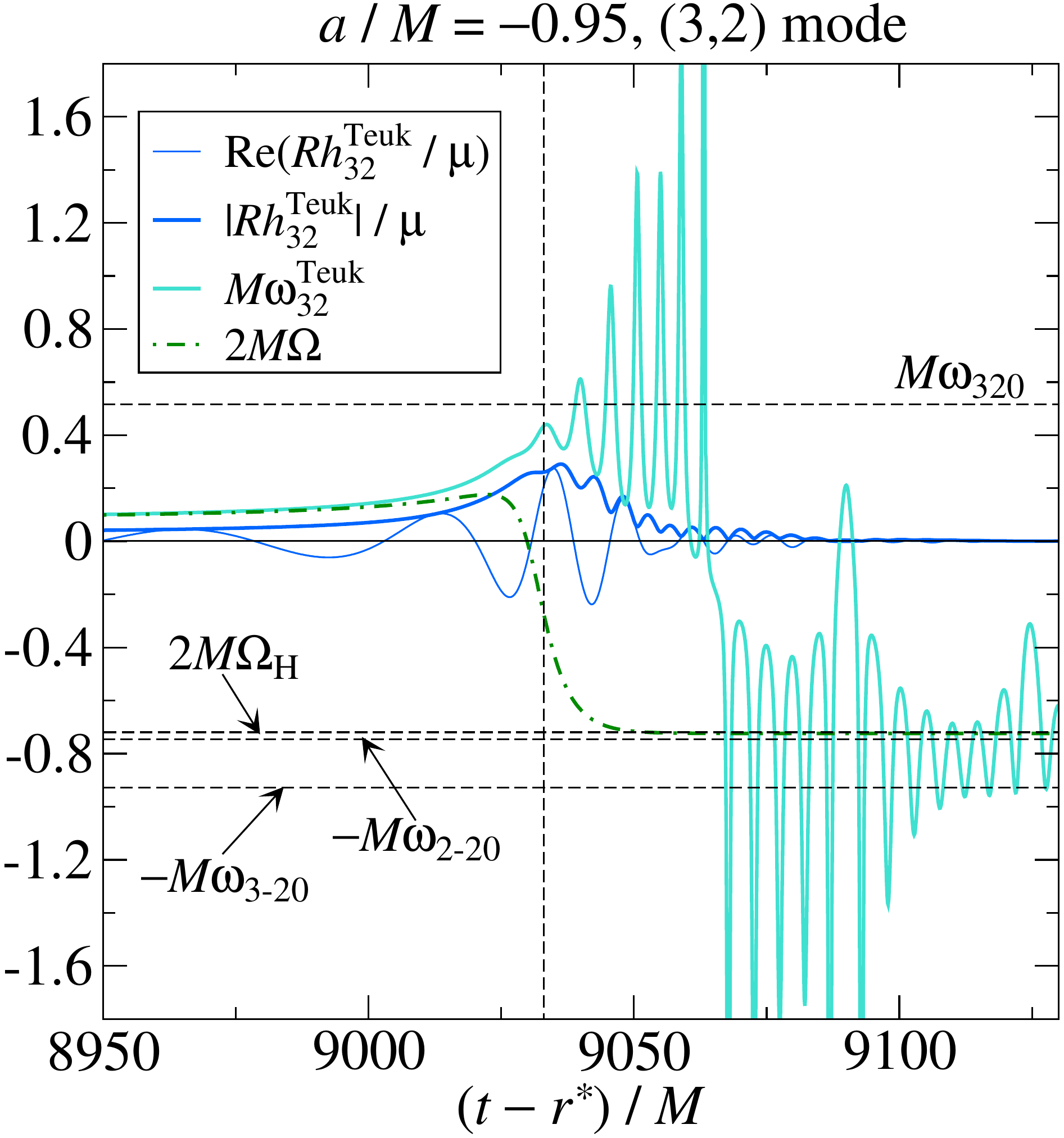}}
 \caption{\label{fig:a_-05_32} Teukolsky $(3,2)$ waveforms for spin $q=-0.5$ (left panel) and $-0.95$ (right panel). For $q=-0.5$, the ringdown waveform contains the modes $(3,-2,0)$ and $(2,2,0)$ besides the usual $(3,2,n)$'s. For the $q=-0.95$ waveform, the GW frequency modulations in the late ringdown are not centered neither about $-\omega_{3-20}$ nor about $-\omega_{2-20}$, and we cannot apply the simple model of Eq.~(\ref{BothRD}). The vertical dashed lines mark $t=t_{\rm match}^{32}$. $R$ is the distance to the source.}
\end{figure*}

QNM mixing manifests itself through striking features in the Teukolsky ringdown waveforms, which are modulated both in amplitude and frequency. To understand the composition of the QNM spectrum of the Teukolsky data, we will study in particular the GW frequency of each mode, defined as $\omega_{\ell m}^{\rm Teuk}\equiv -\Im{(\dot{h}_{\ell m}^{\rm Teuk}/h_{\ell m}^{\rm Teuk})}$, since this quantity is directly related to the frequencies of the most excited QNMs, and is numerically well determined. As an example, Fig.~\ref{fig:Stack32} displays the ringdown $(3,2)$
  mode frequencies for several positive spins, with a common time axis
  rescaled by $2\pi/\omega_{320}$. We observe that different spins have completely different ringdown frequencies; each case has distinct features (spikes, oscillations), occurring with specific periodicities. The averages of the oscillatory features are closer either to $\omega_{320}$ (as one would naively expect) or to $\omega_{220}$, according to the value of $q$. Examples of amplitude modulations can be found in Figs.~\ref{fig:a_-07_22}, \ref{fig:a_-08_21}, and
  \ref{fig:a_-05_32}, which show a few Teukolsky merger-ringdown waveforms
  (solid blue lines), chosen within the large set that we computed for
  this paper.  Among them, the most modulated case is spin $-0.99$ (its $(2,2)$ mode is
  shown in the right panel of Fig.~\ref{fig:a_-07_22}; a similar
  behavior is also present in its higher-order modes). 

In extreme and small mass-ratio binaries, two instances may enhance the excitation and/or mixing of modes other than the $(\ell,m,n)$'s in the ringdown of $(\ell, m)$. On the one hand, for modeling purposes, the strain waveform $h$ is typically decomposed onto $-2$-spin-weighted spherical harmonics $_{-2}Y_{\ell m}$, while the Teukolsky equation is separated using $-2$-spin-weighted spheroidal harmonics $_{-2}S_{\ell m}^{q\omega}$, which depend on the Kerr spin $q$ and the (possibly complex) frequency $\omega$ of the gravitational perturbation. The expansion of the $_{-2}S_{\ell m}^{q \omega}$'s in terms of the $_{-2}Y_{\ell m}$'s can be found (to order $(q M\omega)^{2}$) in Appendix~F of Ref.~\cite{Tagoshi:1996gh}. Using this result, one can derive a formula relating the spherical to the spheroidal waveforms (see, e.g., Eq.~(19) of Ref.~\cite{Pan:2010hz}) and one finds that the spherical mode $h_{\ell m}$ receives contributions from all spheroidal modes with the same $m$, but different $\ell$ (see also Eq.~(38) in Ref.~\cite{Buonanno:2006ui}). Another source of mixing is the orbital motion of the perturbing particle: whenever $q<0$, the orbital frequency switches sign during the plunge, because of frame dragging exerted by the spinning BH; this results in a significant excitation of modes with opposite $m$, but with the same $\ell$. Reference~\cite{Kelly:2012nd} investigated in detail the origin of the mixing in the $(3,2)$ mode of several comparable-mass, nonprecessing BH binaries, and attributed it mostly to angular-basis effects, using $\omega=q_{f} M_{f}\sigma_{320}$.

To understand quantitatively the QNM mixing in our Teukolsky waveforms, we model the ringdown as done in EOB models (i.e., as a linear superposition of overtones of the least-damped QNM), but with the addition of up to 2 further QNMs. While the least-damped mode and its overtones are going to account for the overall shape of the ringdown waveform, the additional QNMs are going to induce the modulations. More explicitly, (except for the $(3,2)$ mode of systems with $q =0.99$) we model the $(\ell,m)$ mode of the ringdown 
waveforms as
\begin{equation}
\begin{split}
h_{\ell m}^{\rm RD} &= \sum_{n=0}^{N-1} A_{\ell m n}e^{-i \sigma_{\ell m n}(t-t_{\rm match}^{\ell m})}\\
&+ \mathcal{S}(t)\left[A_{\ell' m 0} e^{-i \sigma_{\ell' m 0}(t-t_{\rm match}^{\ell m})}+ A_{\ell -m 0} e^{i \sigma_{\ell -m 0}^{*}(t-t_{\rm match}^{\ell m})}\right]\,,\label{BothRD}
\end{split}
\end{equation}
where $t_{\rm match}^{\ell m}$ is the time of merger, $N$ is the number of overtones included, the $A_{\ell m n}$'s are the (constant) coefficients of the overtones, $\mathcal{S}(t)\equiv\left[1 + \tanh{[(t-t_{\rm s})/\tau_{\rm
      s}]}\right]/2$ is a factor introduced to have a smooth switch-on of the interfering QNMs (with $t_{\rm s}$ and $\tau_{\rm s}$ 
optimized mode by mode), and $A_{\ell' m0}$ and $A_{\ell-m0}$ are constants computed from a fit (see below). $A_{\ell' m0}$ and $A_{\ell-m0}$ quantitatively describe the strength of the QNM mixing. Note that $\sigma_{\ell -m n}(M_{f},q_{f})=\sigma_{\ell m n}(M_{f},-q_{f})$. Since overtones with $n>0$ have short decay times with respect to those with $n=0$, Eq.~(\ref{BothRD}) 
is actually dominated by terms with $n=0$ when $t \gg t_{\rm match}^{\ell m}$.

The coefficients $A_{\ell m n}$, $A_{\ell' m 0}$, and $A_{\ell -m 0}$ can be determined from the Teukolsky data as follows. 
Whenever mode mixing is resolved, $A_{\ell' m 0}$ and $A_{\ell -m 0}$ are obtained by fitting the GW frequency $\omega_{\ell m}^{\rm
  RD}= -\Im{(\dot{h}_{\ell m}^{\rm RD}/h_{\ell m}^{\rm RD})}$ to the ringdown Teukolsky GW frequency $\omega_{\ell m}^{\rm Teuk}$,
while setting $A_{\ell m n}=0$ for $n>0$; we choose a fitting window
as wide as possible, but still avoiding any numerical
noise. Once $A_{\ell' m 0}$ and $A_{\ell -m 0}$ are fixed by the fit, the $A_{\ell m n}$'s are calculated via the hybrid
matching procedure detailed in Ref.~\cite{Pan:2011gk}, which consists in a smooth stitching of the ringdown waveform $h_{\ell m}^{\rm RD}$ to the Teukoslky waveform $h_{\ell m}^{\rm Teuk}$ at a time $t_{\rm match}^{\ell m}$.

As in Ref.~\cite{Barausse:2011kb}, we find that, in the test-particle limit, and when the spin is $q\lesssim 0$, some of the physical overtones included in Eq.~(\ref{BothRD}) have frequencies smaller than $\omega_{\ell m}^{\rm Teuk}(t_{\rm match}^{\ell m})$, causing the slope of $\omega_{\ell m}^{\rm RD}$ to be too steep. Therefore, we introduce a pseudo-QNM (i.e., a mode not belonging to the physical QNM spectrum). In the past, pseudo-QNMs were exploited  in comparable-mass EOB models~\cite{Pan:2011gk,Damour:2012ky,Taracchini:2013rva,Pan:2013rra} to reduce the slope of the GW frequency in the transition from plunge to ringdown.

To summarize, the matching procedure has the following tuning parameters: the matching point $t_{\rm match}^{\ell m}$; the size of the
time interval over which one carries out the matching $\Delta t_{\rm
  match}^{\ell m}$; a pseudo-QNM mode with frequency and decay time
$\omega^{\rm pQNM}_{\ell m}$ and $\tau_{\ell m}^{\rm pQNM}$; $t_{\rm
  s}$ and $\tau_{\rm s}$. These tuning parameters are chosen with the goal of minimizing the phase and relative amplitude difference between
$h_{\ell m}^{\rm RD}$ and $h_{\ell m}^{\rm Teuk}$ when $t > t_{\rm match}^{\ell m}$.

Before modeling the entire ringdown waveforms, to better understand how the mixing works, 
let us consider the simple case of just 2 QNMs interfering: let $A_{\ell' m0} = 0$ and $A_{\ell m n} =0$ for $n>0$ 
(i.e., a waveform dominated by the $(\ell,\pm m,0)$ modes). This is similar to what was done in 
Refs.~\cite{Bernuzzi:2010ty,Barausse:2011kb}, where the modulations in the ringdown frequency of the 
numerical modes were fitted with a simple analytical  
formula that accounted for the interference between the $(\ell,\pm m,0)$ QNMs.
 The GW frequency is $\omega_{\ell m}^{\rm RD} = -\Im{(\dot{h}_{\ell m}^{\rm
    RD}/h_{\ell m}^{\rm RD})}$, thus we have (leaving out the factor
  $\mathcal{S}(t)$ for simplicity)
\begin{widetext}
\begin{equation}
\omega_{\ell m}^{\rm RD}=\frac{\omega_{+}-\omega_{-}|\bar{A}|^{2}e^{2(t-t_{\rm match}^{\ell m})\Delta\alpha} + |\bar{A}|e^{\alpha_{+}(t-t_{\rm match}^{\ell m})}\left[\Delta\omega\cos{[\bar{\omega}(t-t_{\rm match}^{\ell m})+\bar{\theta}]} - \Delta\alpha\sin{[\bar{\omega}(t-t_{\rm match}^{\ell m})+\bar{\theta}]}\right]}{1 + |\bar{A}|^{2}e^{2(t-t_{\rm match}^{\ell m})\Delta\alpha}+2|\bar{A}|\cos{[\bar{\omega}(t-t_{\rm match}^{\ell m})+\bar{\theta}]}}\,,\label{omegaMix}
\end{equation}
\end{widetext}
where $\omega_{\pm} \equiv \omega_{\ell \pm m 0}$, $\alpha_{\pm} \equiv 1/\tau_{\ell \pm m 0}$, $\Delta\omega\equiv\omega_{+} - \omega_{-}$, $\Delta\alpha\equiv\alpha_{+}-\alpha_{-}$, $\bar{\omega}\equiv\omega_{+}+\omega_{-}$, and $A_{\ell -m0}/A_{\ell m0}\equiv |\bar{A}|\exp{(i\bar{\theta})}$. Typically, $|\bar{A}| < 1$. Note that Eq.~(19) in Ref.~\cite{Bernuzzi:2010ty} is simpler than our Eq.~(\ref{omegaMix}) above since that paper considered the Schwarzschild case, for which $\sigma_{\ell m n}=\sigma_{\ell-mn}$. Equation~(\ref{omegaMix}) describes a function with exponentially growing oscillations about $\omega_{+}$ when $t_{\rm match}^{\ell m} < t < t_{\rm match}^{\ell m}-\log{|\bar{A}|}/\Delta\alpha \equiv t_{\rm p}$, and with exponentially decreasing oscillations about $-\omega_{-}$ when $ t > t_{\rm p}$; the frequency of the oscillations is $\bar{\omega}$. The point $t_{\rm p}$ marks the transition from oscillations about $\omega_{+}$ to oscillations about $-\omega_{-}$; note that if $t_{\rm p}- t_{\rm match}^{\ell m} \gg 1/\alpha_{+}$ then the transition  occurs in a region where the amplitude is absolutely negligible. Given the size of the numerical errors discussed in Sec.~\ref{sec:TeukolskyCode}, we consider that the ringdown has ended whenever the amplitude drops below $10^{-4}\mu/R$, where $R$ is the distance to the source.

\begin{table*}[!ht]
 \begin{ruledtabular}
    \begin{tabular}{ccccccccc}
     & \multicolumn{2}{c}{$(2,2)$ mode} & \multicolumn{2}{c}{$(3,3)$ mode} & \multicolumn{2}{c}{$(4,4)$ mode} & \multicolumn{2}{c}{$(5,5)$ mode} \\
   $a/M$ & $|A_{2-20}/A_{220}|$ & $\arg{(A_{2-20}/A_{220})}$ & $|A_{3-30}/A_{330}|$ & $\arg{(A_{3-30}/A_{330})}$ & $|A_{4-40}/A_{440}|$ & $\arg{(A_{4-40}/A_{440})}$ & $|A_{5-50}/A_{550}|$ & $\arg{(A_{5-50}/A_{550})}$\\
    \hline \\[-13.5pt]%
 0 & 0.0036 & $-5.70$ & 0.0029 & $-4.07$ & 0.0035 &
   $-5.80$ & 0.0047 & $-7.58$ \\
 $-0.5$ & 0.052 & $-0.24$ & 0.049 & $-1.26$ & 0.056 &
   3.82 & 0.073 & 2.82 \\
 $-0.7$ & 0.12 & $-0.26$ & 0.22 & 4.31 & 0.29 &
   2.86 & 0.39 & 1.60 \\
 $-0.8$ & 0.10 & $-0.41$ & 0.22 & 3.03 & 0.31 &
   1.02 & 0.38 & $-0.81$ \\
 $-0.9$ & 0.28 & $-1.31$ & 0.27 & 1.41 & 0.32 &
   $-1.76$ & 0.46 & $-4.55$\\
  \end{tabular}
\end{ruledtabular}
\caption{Relative amplitude and phase of the QNMs responsible for mixing in modes with $\ell=m$. No QNM mixing is present when the spins are positive. Spins $q=-0.95,-0.99$ cannot be modeled with Eq.~(\ref{omegaMix}) due to the presence of additional interfering QNMs that we are unable to extract, which results in a GW frequency drift at late times (see, for instance, the inset in the right panel of Fig.~\ref{fig:a_-07_22}).\label{tab:Oppositem}}
\end{table*}
\begin{table*}[!ht]\begin{ruledtabular}
     \begin{tabular}{ccccccc}
    & \multicolumn{2}{c}{$(2,1)$ mode} & \multicolumn{4}{c}{$(3,2)$ mode} \\
   $a/M$ & $|A_{2-10}/A_{210}|$ & $\arg{(A_{2-10}/A_{210})}$ & $|A_{220}/A_{320}|$ & $\arg{(A_{220}/A_{320})}$ & $|A_{3-20}/A_{320}|$ & $\arg{(A_{3-20}/A_{320})}$\\
    \hline \\[-13.5pt]%
0.99 & 0.038  & 13.7 & \multicolumn{2}{c}{only $(2,2,n)$ overtones}&&\\
0.95 & 0.030 & 9.82 & 2.91 & 6.44 &  & \\
 0.9 & 0.0025 & 5.61 & 2.11 & 5.99 &  &  \\
 0.8 & 0.0024 & 4.24 & 1.13 & 2.86 &  &  \\
 0.7 & 0.0051 & 3.73 & 0.71 & $-2.10$ &  &  \\
 0.5 & 0.010 & $-0.16$ & 0.34 & $-7.02$ &  &  \\
 0 & 0.069 & $-3.46$ &  &   & 0.010 & $-1.08$ \\
 $-0.5$ & 0.21 & $-5.64$ & 0.13 & $-2.46$ & 0.093 &
   5.26 \\
 $-0.7$ & 0.26 & $-6.58$ & 0.15 & 3.47 & 0.22 &
   10.6 \\
 $-0.8$ & 0.30 & $-7.37$ & 0.16 & 3.91 & 0.23 &
   12.4 \\
 $-0.9$ & 0.32 & $-8.50$ & - & - & - & -\\
 \end{tabular}
    \end{ruledtabular}
\caption{Relative amplitude and phase of the QNMs responsible for mixing in modes with $\ell\neq m$. A blank entry means that that QNM is not excited. A dash indicates that the mode, while present, cannot be reliably fitted. Spins $q=-0.95,-0.99$ and the $(3,2)$ mode of spin $q=-0.9$ cannot be modeled due to the presence of additional interfering QNMs we are unable to extract with the simple model of Eq.~(\ref{BothRD}). The $(3,2)$ mode of spin 0.99 is modeled using only $(2,2,0)$ and its overtones. \label{tab:Oddmodes}}
\end{table*}
\begin{table}
 \begin{ruledtabular}
    \begin{tabular}{cccc}
    $a/M$ & $\Delta t_{\rm match}^{\ell \ell}/M$ & $\Delta t_{\rm match}^{21}/M$ & $\Delta t_{\rm match}^{32}/M$\\
    \hline \\[-12pt]%
    0.99 & 15\,(20)& 15 & 20\\
    0.95 & 13 & 13 & 13\\
    0.9   & 11 & 11 & 11\\
    0.8   & 9 & 9 & 9\\
    0.7   & 7 & 7 & 7\\
    0.5   & 5 & 5 & 5\\
    0      & 5 & 5 & 5 \\
    $-0.5$ & 5 & 3 & 15\\
    $-0.7$ & 5 &3& 15\\
    $-0.8$ & 5& 3 & 20\\
    $-0.9$ & 5 & 3 & -\\
    \end{tabular}
\end{ruledtabular}
\caption{Intervals for ringdown hybrid matching. When $q=0.99$, $\Delta t_{\rm match}^{22}=\Delta t_{\rm match}^{33}=15M$ and $\Delta t_{\rm match}^{44}=\Delta t_{\rm match}^{55}=20M$. The table does not include those spins that we are not able to model.\label{tab:comb}}
\end{table}

In the next two sections we shall discuss how we apply
  Eq.~(\ref{BothRD}) to model the $\ell=m$ and $\ell \neq m$ numerical
  modes, respectively. The main conclusions can be summarized as
follows.  We are able to model the $\ell=m$ modes for any spin $q\geq
-0.9$. The reason why we cannot model spins smaller than $-0.9$ is the
conjectured presence of one or more QNMs that we are unable to
recognize, which manifest themselves in a drift of the Teukolsky GW
frequency at late times (see discussion in Sec.~\ref{sec:even}). We
can model the $(2,1)$ mode for any spin except $q=-0.95,-0.99$ because
of large inaccuracies in capturing modulations in the early
ringdown. Finally, we can model the $(3,2)$ mode for any spin except
$q=-0.9, -0.95$, $-0.99$ because for these very negative spins the
oscillations in the amplitude and GW frequency become very dramatic,
preventing us from reliably fitting the amplitude and phase of all the
modes (see the right panel of Fig.~\ref{fig:a_-05_32}).

Note that we use a unique tuning for the pseudo-QNM, namely $\omega^{\rm pQNM}_{\ell m} = [\omega_{\ell m 0} + \omega_{\ell m}^{\rm Teuk}(t_{\rm match}^{\ell m})]/2$ and $\tau_{\ell m}^{\rm pQNM}=0.2\tau_{\ell m 0}$. When no mode mixing is present, the pseudo-QNM replaces the 8-th physical overtone $(\ell,m,7)$, otherwise it is added to the rest of the mode spectrum. For all the spins that we have been able to model, the matching intervals $\Delta t_{\rm match}^{\ell m}$ are listed in Table~\ref{tab:comb}.

\subsection{$\ell = m$ modes}
\label{sec:even}
For all modes with $\ell = m$, we choose $t_{\rm match}^{\ell m}= t_{\rm peak}^{\Omega}$, which is the time when the orbital frequency $\Omega$ peaks, very close to the light-ring; this choice has the advantage of avoiding the ambiguity of locating the amplitude peak when $q\sim 1$ (see Sec.~\ref{sec:Flat}). For the $(2,2)$ and $(3,3)$ modes we choose $t_{\rm s}=t_{\rm match}^{\ell m} + 20M$ and $\tau_{\rm s}=7.5M$; when $(\ell,m)=(4,4),(5,5)$ we choose instead $t_{\rm s}=t_{\rm match}^{\ell m} + 25M$ and $\tau_{\rm s}=4.5M$.

Let us first consider the dominant $(2,2)$ mode.  For spins $q \gtrsim
0.5$, we find that the ringdown is quite standard, as no appreciable mode mixing is present, and $h_{22}^{\rm Teuk}$
is well described by a linear superposition of overtones of the least-damped mode (i.e.,
Eq.~(\ref{BothRD}) with $A_{\ell' 20},\,A_{2-20}=0$). The matching
interval $\Delta t_{\rm match}^{22}$ varies with $q$ as prescribed in
Table~\ref{tab:comb}. We find that $\Delta t_{\rm match}^{22}$ tends to grow 
towards large, positive spins since the light-ring (i.e., the
matching point) occurs progressively later, during the ringdown, well
past the amplitude peak, in a region where the waveform is rapidly
decaying. For spins $q\leq 0$, we find it necessary to include
the $(2,-2,0)$ mode in the QNM spectrum (i.e., Eq.~(\ref{BothRD}) with $A_{\ell' 20}=0$); this mode has an amplitude
$|A_{2-20}|$ that grows (relative to $|A_{220}|$) as the spin
decreases, which can be understood based on the fact that the portion
of orbit with $\Omega<0$ (due to frame dragging during the plunge)
becomes progressively longer. In Table~\ref{tab:Oppositem} we provide magnitude and phase of $A_{2-20}/A_{220}$, i.e., the ratio of $(2,-2,0)$ relative to the least-damped QNM. The numbers in the table are obtained from a fit of $\omega_{22}^{\rm Teuk}$ using Eq.~(\ref{omegaMix}).

The typical performance of the model is illustrated in the left panel of Fig.~\ref{fig:a_-07_22}, which shows the case with spin $-0.7$. The Teukolsky amplitude (frequency) is plotted in blue (cyan), while the model amplitude (frequency) is plotted in red (orange). We
clearly recognize the growing oscillations of the GW frequency about
$M\omega_{220} \approx 0.31$; by fitting, we find that $|A_{2-20}| /
|A_{220}| \approx 0.12$, so that $t_{\rm p}\approx t_{\rm match}^{\ell m} + 270M$ 
(i.e., in a region where $|Rh_{22}^{\rm Teuk}|/\mu \ll 10^{-4}$). The
waveform $h_{22}^{\rm RD}$ does a good job at capturing the modulations
everywhere, except in the early ringdown ($t_{\rm match}^{\ell m} < t \lesssim
t_{\rm s}$), where the oscillations in $\omega_{22}^{\rm Teuk}$ occur
at a frequency $\bar{\omega} \neq \omega_{220}+\omega_{2-20}$, and
with an amplitude growth whose timescale does not clearly relate to
either $\tau_{220}$ or $\tau_{2-20}$, as one would expect from
Eq.~(\ref{omegaMix}). One limitation inherent to our approach is the
specific form $\mathcal{S}(t)$ of the time
dependence of the coefficients $A_{\ell' m0}$ and $A_{\ell-m0}$, which
may not correctly model the process of excitation (in spite of the two
adjustable parameters $t_{\rm s}$ and $\tau_{\rm s}$). Note that in comparable-mass EOB models the coefficients in front of the QNMs in $h_{\ell m}^{\rm RD}$ have no time dependence. 

For spins
$q<-0.8$, the point $t_{\rm p}$ moves closer to $t_{\rm match}^{\ell m}$, and the performance of the model in the early ringdown (i.e., $t_{\rm match}^{\ell m} < t < t_{\rm p}$) becomes worse; however, Eq.~(\ref{BothRD}) with $A_{\ell' 20} = 0$ can still describe the region $t> t_{\rm p}$ quite accurately. 

We find that the most difficult ringdown waveforms to model are the 
ones with spin $-0.95$ and$-0.99$; the case $q=-0.99$ is shown in the right
panel of Fig.~\ref{fig:a_-07_22}. These are the cases with the longest inversion 
of the trajectory due to frame dragging. We have verified that the numerical
errors during the ringdown are not responsible for creating any of the
modulations. Note that for such extreme (negative) spins we have
$2\Omega_{\rm H}\sim -\omega_{2-20}$. We suspect that the reason why we cannot model spins smaller than $-0.9$ 
is the interference of other QNMs besides those included in Eq.~(\ref{BothRD}), which we are unable to identify; their presence is hinted by the (physical) drift in the GW frequency at late times. This is exemplified in the right panel of Fig.~\ref{fig:a_-07_22}, which refers to the $(2,2)$ mode of spin $-0.99$. The inset therein zooms into the late ringdown, past the point where $\omega_{22}^{\rm Teuk}$ transitions to oscillations about a negative frequency. One can see that the average of the oscillations is not $-\omega_{2-20}$, but instead it slowly asymptotes to that value from above.

As to the other modes with $\ell = m$, they behave similarly to the
$(2,2)$ mode, namely for spins $q > 0$ no significant mode mixing
is present, while for $q \leq 0$ the mode $(\ell,-m,0)$ is
excited. In Table~\ref{tab:Oppositem} we list the
extracted coefficients (relative to the coefficient of the dominant
QNM) of those QNMs that cause amplitude and frequency modulations. Again, the simple ringdown model of Eq.~(\ref{BothRD}) with
$A_{\ell' m0} = 0$ fails to accurately describe the early ringdown for spins
$q<-0.8$, so that we cannot model $q=-0.95,-0.99$.

We have also tried to look for contributions from the horizon modes suggested by
Refs.~\cite{Mino:2008at,Zimmerman:2011dx}, whose frequency is 
$m\Omega_{\rm H}$, but their decay time $r_{+}/(2\sqrt{1-q^{2}})$ is not
compatible with any of the timescales present in the Teukolsky data, 
and we did not observe their presence in the numerical waveforms.

\subsection{$\ell \neq m$ modes}
\label{sec:odd}

We find that the $(2,1)$ mode shows mode mixing all across the physical spin range:
the $(2,-1,0)$ component can be excited also for $q>0$, although to a limited extent. Explicitly, we model its ringdown via Eq.~(\ref{BothRD}), setting $A_{\ell' 1 0}=0$. If $q>0$ we can choose $t_{\rm match}^{\ell m} = t_{\rm peak}^{\Omega}$, $t_{\rm
  s} = t_{\rm match}^{21} + 15M$, $\tau_{\rm s}=7.5M$. For positive spins, the amplitude of $(2,-1,0)$ (shown in Table~\ref{tab:Oddmodes}) turns out to be rather small ($|A_{2-10}/A_{210}|\sim 10^{-3}\mbox{--}10^{-2}$). The model performs very well in this region. 

Starting from the nonspinning case, and for smaller spins, $t_{\rm peak}^{\Omega}$ occurs quite early with respect to the beginning of the $(2,1)$ ringdown, therefore we find it necessary to modify our matching prescriptions. The option of choosing $t_{\rm peak}^{21}$ is certainly viable for $q=0$. However, when $q<0$, the onset of
mode mixing is quite prompt, so that even the amplitude peak itself is
affected by it. To illustrate this point, in the left panel of Fig.~\ref{fig:a_-08_21} we plot the $(2,1)$ merger-ringdown waveform for spin $-0.8$; the left half of the amplitude peak is standard, whereas the right half is modulated by the QNM mixing (featuring several bumps). We observe that the amplitude oscillations begin at the turning point of the particle's azimuthal motion (i.e., when $\Omega$ vanishes); thus, we choose this as our $t_{\rm match}^{\ell m}$ for negative spins. We also choose $t_{\rm s} = t_{\rm match}^{\ell m}+10M$ and $\tau_{s}=7.5M$. The correct modeling of the amplitude modulations critically depends on the prescriptions used for the matching, in
particular $\Delta t_{\rm match}^{21}$, which can be found in Table~\ref{tab:comb}. The model performs quite well for spins as small as $-0.9$, except for the first couple of oscillations induced by QNM mixing, as can be seen in the left panel Fig.~\ref{fig:a_-08_21}, mainly due to $\mathcal{S}(t)$.
In spite of the different ringdown prescriptions used in the positive versus negative spin regime, we can see from Table~\ref{tab:Oddmodes} that $|A_{2-10}/A_{210}|$ and $\arg{(A_{2-10}/A_{210})}$ are well-behaved functions of spin. Spins $-0.95$ and $-0.99$ cannot be modeled accurately, the issue being the early ringdown (i.e., $t_{\rm match}^{21}<t<t_{\rm p}$), whose modulations become rather extreme, and are not captured by $\mathcal{S}(t)$. The late ringdown (i.e., $t>t_{\rm p}$) follows instead well our model.

The more challenging mode to model is the $(3,2)$. For $q>0$ 
we use $t_{\rm match}^{\ell m}=t_{\rm peak}^{\Omega}$, $t_{\rm s}=t_{\rm match}^{32}+2.5M$, and $\tau_{\rm s}=10M$. As already seen in
Fig.~\ref{fig:Stack32}, when $q \gtrsim 0.7$ the QNM mixing induces a transition of the
average (final) ringdown frequency from the expected least-damped mode
frequency $\omega_{320}$ to $\omega_{220}$. Note
how the case $q=0.7$ (fifth panel of Fig.~\ref{fig:Stack32}) sits at
the transition between the two regimes, featuring wide frequency
oscillations around both $\omega_{320}$ and $\omega_{220}$. The case $q=0.99$ stands out, since its
ringdown can be described by the $(2,2)$-mode spectrum (i.e.,
$h_{32}^{\rm RD}=\sum_{n}A_{22n}\exp{[-i \sigma_{22n}(t-t_{\rm match}^{\ell m})]}$
is a good model for $h_{32}^{\rm Teuk}$). This happens because there are 
no significant mode-mixing modulations (see the first panel of
  Fig.~\ref{fig:Stack32}) and the asymptotic GW frequency is
$\omega_{32}^{\rm Teuk}(t\to \infty) = \omega_{220}$. In the range $0 < q \leq 0.95$, instead, we model the ringdown via Eq.~(\ref{BothRD}), setting $A_{3-20}=0$ and $\ell'=2$, i.e., the QNM spectrum is that of the $(3,2)$ mode with interference from $(2,2,0)$. When
$0.8\lesssim q \lesssim 0.95$, the ringdown displays large features,
with a GW frequency oscillating about $\omega_{220}$ (see second
  to fourth panel of Fig.~\ref{fig:Stack32}); this means that $(2,2,0)$ is more excited than the least-damped mode $(3,2,0)$, which is confirmed by our fits, as $|A_{220}/A_{320}|>1$ (see Table~\ref{tab:Oddmodes}). The right panel of
  Fig.~\ref{fig:a_-08_21} shows the good agreement of the model to the
  Teukolsky data for $q=0.9$. Notice how the matching point lies in a
  region where the amplitude has already started to drop, quite a bit
  later than the peak. As already discussed, the case with $q=0.7$
represents a sort of threshold, in that its GW frequency oscillates
about $\omega_{320}$ in the early ringdown and then about
$\omega_{220}$ in the late ringdown (see the fifth panel of
  Fig.~\ref{fig:Stack32}). 

Similarly to the $(2,1)$ mode, when $q\leq 0$, $t_{\rm peak}^{\Omega}$ occurs quite early; when $q<0$, the $(3,2)$ amplitude peak is modulated by the mode
mixing, but now the turning point of the particle happens somewhat earlier relative to
it. Therefore, when $q\leq0$, we choose the matching point in the ``middle'' of the amplitude peak, where $\partial_{t}^{3}|h_{32}^{\rm Teuk}|=0$; we also choose $t_{s}=t_{\rm match}^{32}+10M$ and $\tau_{s}=7.5M$. In terms of QNM spectrum, as happens for all the modes we studied, for $q\leq0$
the mode with opposite $m$ is excited (i.e., $(3,-2,0)$). However, the
$(2,2,0)$ mode can still be extracted from the $(3,2)$ waveforms for
spins as small as $-0.8 \leq q < 0$. Here we use Eq.~(\ref{BothRD}) with $A_{3-20},\,A_{220}\neq0$. The function that we fit to the Teukolsky data is simply the generalization of Eq.~(\ref{omegaMix}) to three interfering QNMs. 
The extracted coefficients are found in Table~\ref{tab:Oddmodes}. An example of this regime
  is shown in the left panel of Fig.~\ref{fig:a_-05_32}, for spin
  $-0.7$; one can notice two effects in $\omega_{32}^{\rm Teuk}$, the
  high-frequency modulations due to the interference of $(3,-2,0)$,
  and the low-frequency ones due to the interference of $(2,2,0)$. It
  is also possible to appreciate how well the model (red and orange
  lines) can capture all these features. For spins $q\leq-0.9$, the
waveforms asymptote to a frequency lying between $-\omega_{2-20}$ and $-\omega_{3-20}$, and we cannot extract the coefficients because $\omega^{\rm Teuk}_{32}$ has a very irregular behavior and we find it hard to determine the appropriate 
fitting window. This problematic regime is depicted in the right panel of Fig.~\ref{fig:a_-05_32}, where we plot the $(3,2)$ mode of spin $-0.95$.

\section{Considerations on the modeling of comparable-mass binary systems}
\label{sec:EOBmodeling}

In this section we explain how the findings of Secs.~\ref{sec:Flat} and \ref{sec:ModeMix} can help EOB waveform modeling.

The EOB approach employs factorized analytical (multipolar) waveforms that 
resum the circular PN formulae, while incorporating strong-field and non-circular
effects~\cite{Damour:2007xr,Damour:2008gu,Pan:2010hz,Nagar:2011aa,Taracchini:2013wfa}~\footnote{Reference~\cite{Taracchini:2013wfa} (see
Appendices~C and D) computed mode-by-mode amplitude fits of the
Teukolsky modes generated by a frequency-domain code, which assumed
circular orbits in Kerr. High, unknown PN terms in the factorized
waveforms were fitted up to the ISCO, for both ingoing and outgoing
radiation.}. An
example of strong-field feature is the divergence of the factorized modes at 
the light-ring for circular orbits through the ``source'' term proportional to the binding energy (angular
momentum) for $\ell = m$ ($\ell \neq m$) modes. Deviations from circularity are
modeled in the EOB waveforms through a phenomenological non-quasicircular (NQC) factor
that reshapes the EOB factorized waveforms during plunge and around merger in order to
better match the numerical waveforms (computed either with
numerical-relativity or Teukolsky-equation codes). The NQC factor is determined
once the numerical ``input values'' (i.e., the amplitude $|h_{\ell m}^{\rm num}|$,
the slope $\partial_{t}|h_{\ell m}^{\rm num}|$, the curvature
$\partial^{2}_{t}|h_{\ell m}^{\rm num}|$, the frequency $\omega_{\ell
  m}^{\rm num}$, the slope of the frequency $\partial_{t}\omega_{\ell
  m}^{\rm num}$) are prescribed. Typically, the input values are read
off at the peak of the numerical waveforms, and, on the EOB side, they
are enforced at a specific time relative to the peak of the orbital
frequency (which occurs at $t=t_{\rm peak}^{\Omega}$). That same time is used as the
attachment point for the ringdown waveform (for more details, see Sec.~IV of
Ref.~\cite{Barausse:2011kb}).

As discussed in Sec.~\ref{sec:Flat}, the very circular character of the 
Teukolsky waveforms when $q \to 1$ is very appealing from the point of view of the modeling, since the
EOB factorized modes (without NQC corrections) are built under the
assumption of quasicircular adiabatic motion. However, highly spinning
systems are also very relativistic, and current PN waveforms (on which the factorized ones
are based) are not accurate enough for such regimes already hundreds
of cycles before merger. As already pointed out in
Refs.~\cite{Pan:2010hz,Barausse:2011kb} by comparisons with frequency-domain Teukolsky
waveforms, due to the lack of enough PN knowledge in the test-particle
limit, the amplitude of the factorized waveforms performs poorly even
before the ISCO for large spins, implying also inaccurate multipolar
fluxes. While we were finalizing this paper, Ref.~\cite{Shah:2014tka} was posted; the author computed the energy fluxes for a particle in circular, equatorial orbit in Kerr spacetime up to 20PN order. In spite of the high PN order of the calculation, the relative accuracy of the analytical flux (when compared to numerical Teukolsky data) is within $10^{-3}$ only down to $2.97 r_{\rm ISCO}$ for spin 0.9, i.e., for an orbital speed around 0.37 (to be compared with $v_{\rm ISCO}\approx 0.61$). As we shall see in Sec.~\ref{sec:CompareWaveforms}, even modeling errors as small as $10^{-3}$ at the ISCO may result in large dephasings once the analytical fluxes are employed in time evolutions.

Moreover, as originally found in Ref.~\cite{Barausse:2011kb}, the larger the spin, the earlier the $(2,2)$ mode
peaks with respect to $t_{\rm peak}^{\Omega}$: when $q \geq 0.9$, the
peak occurs before the ISCO, during the inspiral phase, where the
radial motion is absolutely negligible, as we discussed in Sec.~\ref{sec:Flat}. 
As a consequence, when calculated at the amplitude peak, the NQC functions are heavily
suppressed for large and positive spins, because they are proportional
to $p_{r^{*}} \propto \dot{r}$ (see Fig.~\ref{fig:PlungeVel}), and cannot
help correcting the waveform. One could see what can be gained by applying the factorized resummation procedure to the PN-expanded fluxes of Ref.~\cite{Shah:2014tka}, or keep the current factorized flux while including the fits of
Ref.~\cite{Taracchini:2013wfa}, and obtain EOB amplitudes in greater
agreement with the numerical ones without any need for NQC
corrections. Note that the fits of Ref.~\cite{Taracchini:2013wfa} were computed up to the ISCO. Hence, after the
peak, when the amplitude is falling off, the EOB waveform with fits
can still differ from the Teukolsky one. However, applying an NQC correction at
that late stage could be a viable option.

Furthermore, if we followed the standard EOB prescription of 
attaching the ringdown waveform at $t_{\rm peak}^{\ell m}$, we would not be able to successfully model the Teukolsky 
waveform, because its ringdown sets in at times which are rather close to the peak of the orbital frequency at time $t_{\rm peak}^{\Omega}$, 
while $t_{\rm peak}^{\ell m} \ll t_{\rm peak}^{\Omega}$. As we shall see below, 
to overcome this issue, we suggest a new prescription for the matching point of the ringdown in the EOB approach 
for small mass-ratios and large spins.

These findings for large spins were effectively exploited in the construction of the
EOB model of Ref.~\cite{Taracchini:2013rva}, which extended the model
of Ref.~\cite{Taracchini:2012ig} to generic mass ratios and spins;
only the dominant $(2,2)$ mode was considered. The model was
calibrated to 38 numerical-relativity nonprecessing waveforms produced
by the SXS Collaboration~\cite{Mroue:2013xna,Mroue:2012kv,Hemberger:2012jz,Hemberger:2013hsa},
spanning mass ratios from 1 to 8, spin magnitudes up to 0.98, and with
40 to 60 GW cycles. By construction, any EOB model incorporates the
test-particle limit, since the whole formalism is based on a
deformation of the Kerr spacetime~\footnote{The deformation parameter is 
the symmetric mass ratio $m_1 m_2/(m_1+m_2)^2$, $m_1$ and $m_2$ being the BH masses.}.
As explained above, the merger waveform critically depends on the information from numerical-relativity 
waveforms, in the form of input values. Since numerical-relativity
simulations are still unable to explore the small mass-ratio
limit~\footnote{A roadmap for future, challenging numerical-relativity
  simulations is outlined in the first paper of the NRAR
  Collaboration~\cite{Hinder:2013oqa}.}, the Teukolsky waveforms are
extremely valuable in bridging the gap between mass ratio $\sim1/10$
and $\sim 1/1000$. 

The prototype nonprecessing, spinning EOB model 
of Ref.~\cite{Taracchini:2012ig} (which could cover spins only up to 0.6) introduced, for the first time, a spin-dependent (negative) time
delay $\Delta t^{22}_{\rm peak}$ between $t^{\Omega}_{\rm peak}$ and
the peak of $|h_{22}|$, which was inspired by the time delay seen in the Teukolsky
data of Ref.~\cite{Barausse:2011kb}. Such time delay had already been found in Ref.~\cite{Bernuzzi:2010xj}  
for the (2,2) mode in nonspinning binaries with small mass ratio,  
but because the time delay in the nonspinning case is quite small, it was not needed when modeling the 
(2,2) mode of nonspinning, comparable-mass systems~\cite{Pan:2011gk}. Furthermore, Ref.~\cite{Taracchini:2012ig} fixed the small mass-ratio
limit of $\omega_{22}^{\rm num}$ and $\partial_{t}\omega_{22}^{\rm
  num}$ based on the Teukolsky waveforms of Ref.~\cite{Barausse:2011kb}. 
In the same spirit of Ref.~\cite{Taracchini:2012ig}, some of us used 
the additional information on the test-particle limit provided in this paper 
(in particular, the behavior of the Teukolsky waveforms beyond
spin 0.8) to extend the nonprecessing EOB model to any spin and mass ratio~\cite{Taracchini:2013rva}. 

First, we built a time-delay function $\Delta t^{22}_{\rm peak}$ that, in the
small mass-ratio limit, decreases with spin beyond 0.8; this
guarantees that the ringdown starts close to $t^\Omega_{\rm peak}$ and that the 
NQC equations are always enforced in a region with significant radial motion (at time $t_{\rm peak}^{\Omega}+\Delta
t_{\rm peak}^{22}$), as opposed to the extremely circular region
around the amplitude peak (at time $t_{\rm peak}^{22}$). As an example, in Fig.~\ref{fig:Full22_099} we indicate with a vertical green line where the point $t_{\rm peak}^{\Omega}+\Delta
t_{\rm peak}^{22}$ occurs for such time-delay function when the mass ratio is $1/1000$ and the spin is 0.99: the point safely lies well after the ISCO, close to the light-ring. Remember that the analysis of Sec.~\ref{sec:even} has shown that for $(2,2)$ modes and large spin one can reliably attach the ringdown waveform at the light-ring. 

Second, we
built piecewise continuous fitting functions for the input values
along the spin dimension~\footnote{Note that, in both EOB models of
  Refs.~\cite{Taracchini:2012ig} and \cite{Taracchini:2013rva}, the
  input values are functions of only two parameters: the symmetric
  mass ratio and an effective spin (see definition in Eq.~(32) of
  Ref.~\cite{Taracchini:2012ig}).} such that, beyond spin 0.8 and for mass ratio smaller than $\sim 1/100$, they
approach $|h_{22}|$,  $\partial_{t}|h_{22}|$, $\partial^{2}_{t}|h_{22}|$, $\omega_{22}$,
$\partial_{t}\omega_{22}$ of the EOB factorized waveform itself
(without any spinning NQC correction), evaluated at time $t_{\rm
  peak}^{\Omega}+\Delta t_{\rm peak}^{22}$. This entails that, beyond
spin 0.8 and for mass ratio $1/1000$, the EOB model will not agree
too well with the Teukolsky waveforms produced with the numerical 
flux, as in this paper. This is mainly a consequence of the limitation of the
current factorized waveforms that we discussed above (especially as far as the
amplitude is concerned). Imposing the exact Teukolsky input values
beyond spin 0.8 at mass ratio $1/1000$ would result in unwanted features (such as bumps in
the inspiral amplitude), because the NQC corrections act only over
short time intervals, while the factorized waveforms are discrepant
over much longer spans for this corner of the parameter space. This 
limitation will be overcome once the current factorized waveforms 
are improved.

Third, in the model of Ref.~\cite{Taracchini:2013rva}, the Teukolsky
waveforms were also exploited to establish robust ringdown prescriptions in
the small mass-ratio limit, especially for binaries with large spins. Indeed, 
we found 
it necessary to introduce mass-ratio and spin dependence
in the ringdown tuning parameters (i.e., the size of the matching interval,
frequency and decay time of the pseudo-QNMs). 

Finally, in Appendix~\ref{app:InputValues} we provide input values
measured from the Teukolsky waveforms of this paper, as well as the measured
time delay $\Delta t_{\rm peak}^{\ell m}$, as functions of the spin. This 
data can be used for future, improved versions of the EOB model.

\section{The comparable-mass effective-one-body model in the test-particle limit}
\label{sec:CompareWaveforms}
\begin{figure*}[!ht]
  \begin{center}
    \includegraphics*[width=\textwidth]{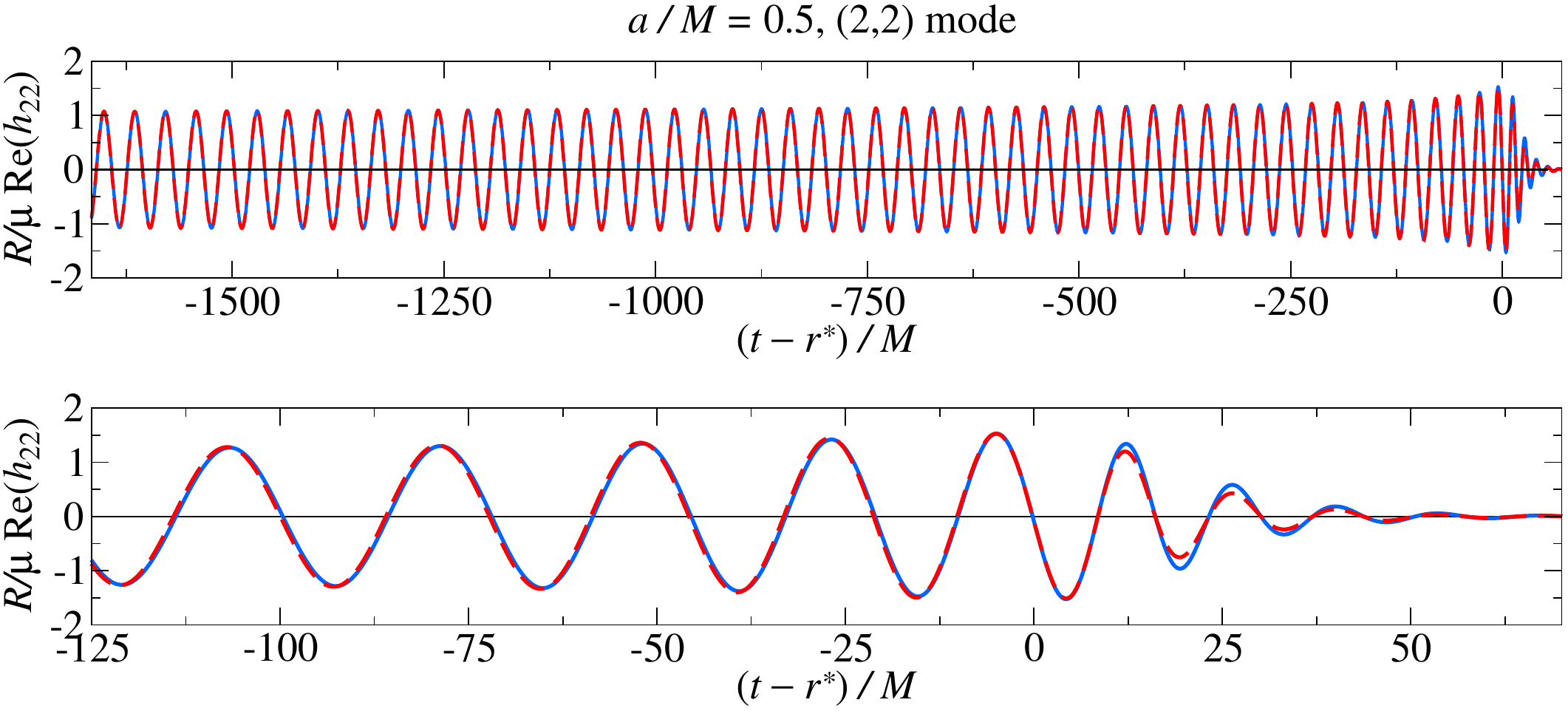}
      \caption{\label{fig:SEOBNRv2_05} For spin 0.5, comparison between Teukolsky $(2,2)$ mode waveform (solid blue lines) and the EOB model of Ref.~\cite{Taracchini:2012ig} evaluated in the test-particle limit (dashed red lines). The Teukolsky waveform is evaluated along the EOB trajectory. The waveforms are aligned at their amplitude peak, which corresponds to 0 retarded time; 50 GW cycles before the peak are shown. $R$ is the distance to the source.}
  \end{center}
\end{figure*}
\begin{figure*}[!ht]
  \begin{center}
    \includegraphics*[width=\textwidth]{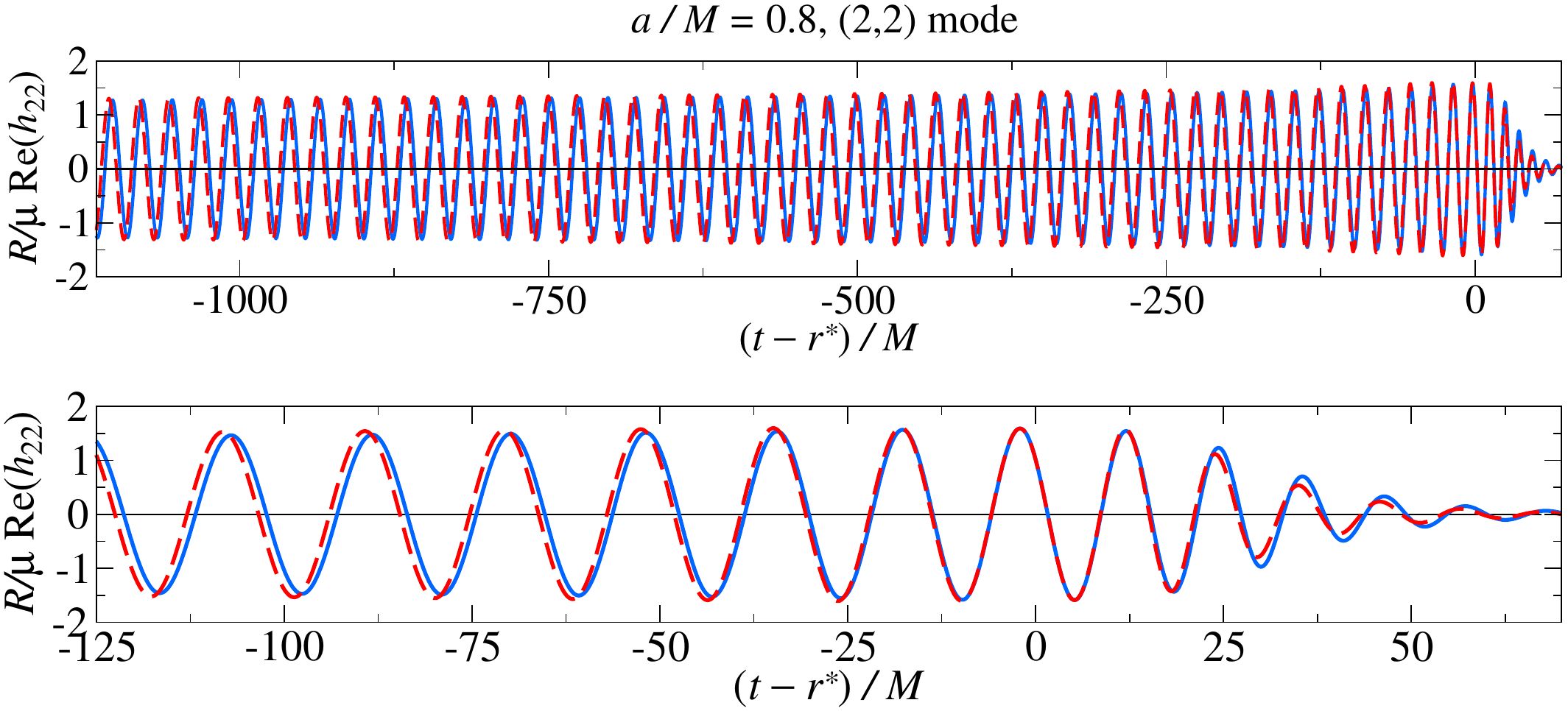}
      \caption{\label{fig:SEOBNRv2_08} Same as Figure~\ref{fig:SEOBNRv2_05}, but for spin 0.8.}
  \end{center}
\end{figure*}
In this section we compare the comparable-mass EOB model of Ref.~\cite{Taracchini:2012ig} to the 
numerical waveforms computed via the Teukolsky formalism in the test-particle limit. Before discussing
the waveforms, we have to point out that the orbital dynamics
generated by the EOB model in this section is quite different from that generated
following the prescriptions of Sec.~\ref{sec:orbdyn}. In fact, as already discussed, 
the EOB energy flux used in Ref.~\cite{Taracchini:2012ig}, which was based on Refs.~\cite{Damour:2008gu,Pan:2010hz} 
and used all the PN corrections available at the time of publication, has several shortcomings in the test-particle limit. 

First, the EOB energy flux used in Ref.~\cite{Taracchini:2012ig} does not account for the ingoing portion of the GW
flux. Horizon absorption has the largest effect for nearly extremal
positive spins, thanks to the slower rate of energy loss, due to
superradiance. Note that the relative sign between ingoing and
outgoing fluxes changes when the orbital frequency crosses the horizon
frequency. When $\Omega \leq \Omega_{\rm H}$ and $q>0$,
 the ingoing fraction subtracts from the outgoing
flux; otherwise, the absorption flux adds to the outgoing flux. For instance, when the spin is 0 (0.99), the absorption flux
increases (decreases) dissipation by $\sim 0.3\%$ ($\sim 9\%$)
for a particle orbiting at the ISCO (see Fig.~1 of
Ref.~\cite{Taracchini:2013wfa}). References~\cite{Hughes:1999bq,Hughes:2000ssa}
found that in the nearly extremal case $q=0.998$ the inspiral up to
the ISCO can be longer by $\sim5\%$ at low inclinations, depending on
whether the ingoing flux is included or not. A study extending up to
merger was done in the Schwarzschild case by
Ref.~\cite{Bernuzzi:2012ku}, which considered an EOB evolution
including the model absorption flux of Ref.~\cite{Nagar:2011aa}; when
the symmetric mass ratio is $10^{-3}$, they found a dephasing of
1.6~rads for the (2,2) mode waveform at merger over an entire
evolution of about 41 orbital cycles. As to the spinning case,
Ref.~\cite{Yunes:2010zj} included the spinning horizon flux in an EOB
model, using the Taylor-expanded expressions of
Refs.~\cite{Tagoshi:1997jy, Mino:1997bx}; the inclusion of absorption
turned out to be important to obtain good agreement with the full
Teukolsky flux, at least up to the ISCO. When modeling spinning
binaries, one should bear in mind that the spin changes the PN order
(with respect to the leading order flux at infinity) at which
absorption enters in the energy flux: while this effect enters at 4PN
order for Schwarzschild BHs, it enters at 2.5PN order for nonzero
spin. 

To confirm the impact of neglecting the ingoing flux, we evolve
trajectories with either the total or only the outgoing Teukolsky
flux, relying again on the data of Ref.~\cite{Taracchini:2013wfa}. We
consider $(2,2)$ waveforms that begin 100 GW cycles before the
ISCO. For comparison, we align their phases both at low frequency
(over the first 10 GW cycles) and at high frequency~\footnote{Note that
  aligning the waveforms at the amplitude peak is not an option, given
  their extreme flatness when $q=0.99$.} (over the 10 GW cycles
following the ISCO), and then measure the phase difference either
during ringdown (for the low frequency alignment) or at the beginning
of the waveform (for the high frequency alignment), using the case
with the total flux as fiducial. After the low frequency alignment, we
find that for spin 0 (0.99) the horizon absorption induces a dephasing
of about $-2$ ($+23$) rads. After the high frequency alignment, we
find that for spin 0 (0.99) the horizon absorption induces a dephasing
of about $-0.1$ ($+8$) rads. The different sign in the dephasings for
spin 0 and 0.99 reflects the fact that for $q\leq0$ the ingoing flux
increases the rate of dissipation (thus hastening the coalescence),
while for $q>0$ superradiance extracts energy from the rotation of the
massive BH and transfers it into the orbital motion (thus delaying the
coalescence). These effects can play a major role for space-based GW
detectors, whose integration time will have to be of the order of
$10^{6}$ GW cycles (or more) to achieve detection~\cite{AmaroSeoane:2007aw},
hence requiring very long and accurate GW templates. 

In principle,
horizon absorption may also alter the merger waveform, which
constitutes a numerical input for the EOB model via the NQC procedure
outlined in Sec.~\ref{sec:EOBmodeling}. For $q=0$ we compute the
$(2,2)$ mode input values $|h_{22}^{\rm Teuk}|$, 
$\partial_{t}^{2}|h_{22}^{\rm Teuk}|$, $\omega_{22}^{\rm Teuk}$,
$\partial_{t}\omega_{22}^{\rm Teuk}$ at $t_{\rm peak}^{22}$ (here, of course, $\partial_{t}|h_{22}^{\rm Teuk}|=0$), while for
$q=0.99$, due to the flatness of the amplitude and the lack of an
orbital frequency peak, we compute them at the ISCO~\footnote{For
  $q=0.99$, the ISCO is only $0.3M$ away from the horizon in the
  radial coordinate. See also Fig.~\ref{fig:Full22_099} for a more precise idea in the case of the $(2,2)$ mode.}. For spin 0 (0.99), the relative difference
induced by horizon absorption on the four input values is
respectively: $0.0014\%$ ($0.17\%$), $0.50\%$ ($5.8\%$), $0.082\%$
($0.29\%$), and $0.091\%$ ($5.2\%$). Similar results apply to
higher-order modes. The larger discrepancies can be seen on the
curvature and on the slope of the GW frequency, but the NQC procedure
is only mildly sensitive to these two quantities, as the most
important features to reproduce are the amplitude and the GW
frequency, which means that the horizon absorption does not impact the
merger waveform significantly.

Second, as compared to the total outgoing Teukolsky flux, the current EOB energy flux does not account for modes with $\ell > 8$. We can quantitatively assess this truncation error in the frequency domain by using the multipolar components of the Teukolsky fluxes computed in Ref.~\cite{Taracchini:2013wfa}. We find that, for a particle orbiting at the ISCO, the fractional contribution to the total outgoing flux coming from modes beyond $\ell = 8$ varies between $10^{-5}$ for $q=-0.99$ and $3\times10^{-3}$ for $q=0.99$. The growing relevance of higher modes with spin is consistent with the trend that one sees when studying the amplitude hierarchy between the dominant $(2,2)$ mode and higher modes~\cite{Barausse:2011kb}. For spins $q=0,0.99$, we compute the Teukolsky waveforms along trajectories sourced by Teukolsky flux modes only up to $\ell =8$, and compare them to the waveforms generated using the total outgoing flux (taken as fiducial). We measure the dephasings with the same approach discussed above when studying the effect of horizon absorption. After the low frequency alignment, we find that for spin 0 (0.99) the higher-$\ell$ modes induce a dephasing of about $-0.3$ ($-7.5$) rads. After the high frequency alignment, we find that for spin 0 (0.99) the higher-$\ell$ modes induce a dephasing of about $-0.015$ ($-3$) rads. The negative signs indicate that, obviously, whenever we neglect $\ell > 8$ modes the rate of dissipation is lower, hence the coalescence occurs later. These phase differences are less dramatic than those seen when neglecting the ingoing flux. Nonetheless, they are relevant for the purpose of generating templates for extreme and small mass-ratio inspirals.

Third, as discussed in Sec.~\ref{sec:EOBmodeling}, the modeling error on the amplitude of the individual factorized modes with $\ell \leq8$ can be significant even before the ISCO for large spins: a more quantitative assessment of the disagreement with numerical amplitudes can be found in Ref.~\cite{Pan:2010hz}. The origin of the poor performance lies in the limited PN knowledge, since for large spins the ISCO moves to a more relativistic regime: $v_{\rm ISCO}\approx 0.41$ when $q=0$, while $v_{\rm ISCO} \approx 0.79$ when $q=1$. Again, one could include the amplitude fits of Ref.~\cite{Taracchini:2013wfa} or apply the factorized resummation to the analytical energy flux of Ref.~\cite{Shah:2014tka}, and recalibrate the comparable-mass model to numerical-relativity simulations.

We now move on to discuss the waveforms. We evaluate the comparable-mass EOB model of Ref.~\cite{Taracchini:2012ig} in the test-particle limit by setting the symmetric mass-ratio $\mu/M$ to zero everywhere in the model, except in the leading term of the GW flux, where we set it to $10^{-3}$; this choice is consistent with the prescriptions of Sec.~\ref{sec:orbdyn} for building orbital evolutions with the Teukolsky fluxes. The GW flux of the model is a sum of time derivatives of multipolar modes up to $\ell = 8$, according to Eq.~(13) of Ref.~\cite{Taracchini:2012ig}. All the modes are the $\rho$-resummed factorized ones of Ref.~\cite{Pan:2010hz}, except those with $\ell\leq4$ and odd $m$, which instead follow the prescription given in appendix A of Ref.~\cite{Taracchini:2012ig}; test-particle limit nonspinning effects are included up to 5.5PN order (beyond the leading order), while spinning effects are included up to 4PN order (beyond the leading order). Here we are not interested in testing the EOB orbital dynamics, but we rather want to focus on the waveforms, therefore the Teukolsky waveforms are calculated along the EOB trajectories. The same approach was adopted in Ref.~\cite{Barausse:2011kb} for the case with spin 0. For spins as large as $q\sim 0.5$, the EOB waveforms are in good agreement with the numerical waveforms. In Fig.~\ref{fig:SEOBNRv2_05}, for $q=0.5$, we align EOB and Teukolsky $(2,2)$ mode waveforms at the amplitude peak; we find a dephasing within 0.1 rads and a relative amplitude error which is negligible everywhere except during ringdown (where it is around 30\%). For larger spins, however, a large discrepancy in the amplitude shows up well before merger. In Fig.~\ref{fig:SEOBNRv2_08}, for $q=0.8$, we find an amplitude error around 5\% during the late inspiral; the dephasing is quite large too, reaching about 0.8 rads 50 GW cycles before merger, and growing as one moves to lower frequencies.

\section{Conclusions}
\label{sec:conclusions}

Using the Teukolsky equation in the time domain, we have computed inspiral-merger-ringdown waveforms produced by the inspiraling motion of a nonspinning test particle in the equatorial plane of a Kerr BH with dimensionless spin $-0.99 \leq q \leq 0.99$, thus extending work done in Ref.~\cite{Barausse:2011kb}. The trajectory of the particle has been obtained from the geodesic equation, subject to a radiation-reaction force that is proportional to the total energy flux in GWs. We have used the GW fluxes computed for circular orbits down to the light-ring with a frequency-domain Teukolsky code~\cite{Taracchini:2013wfa}. We have computed the dominant and leading subdominant modes of the radiation: $(2,2)$, $(2,1)$, $(3,3)$, $(3,2)$, $(4,4)$, and $(5,5)$.

In Sec.~\ref{sec:Flat}, we have pointed out the simplicity of the waveforms emitted by systems with large, positive spins, in spite of the highly relativistic regime probed by the inspiraling orbital trajectories. The main feature of the mode amplitudes is their flattening towards the ISCO and 
during the plunge as the spin grows (see Fig.~\ref{fig:Flat22LargeSpin}). We have given an explanation of this phenomenon in terms of the ratio between the orbital and the radiation-reaction timescales. On the one hand, as $q\to1$ the total (i.e., ingoing $+$ outgoing) GW flux tends to decrease, partly thanks to the extraction of energy from the rotation of the Kerr BH via superradiance. On the other hand, as $q\to1$ the horizon (i.e., the final point of the orbital evolution) moves to smaller radii, which implies higher orbital frequencies accessible to the inspiraling particle. This results in a significant increase in the number of orbits per unit frequency as $q\to1$; the orbital motion becomes extremely circular, and highly relativistic. 

In Sec.~\ref{sec:ModeMix}, we have systematically studied the ringdown stage, whose waveforms display complicated amplitude and frequency modulations due to the interference of QNMs. In the comparable-mass range, with the notable exception of the $(3,2)$ mode, the $(\ell,m)$ modes of nonprecessing BH binaries can be successfully modeled by the linear superposition of overtones of the least-damped QNM, i.e., $(\ell,m,n)$, with $n=0,1,\cdots$~\cite{Buonanno:2000ef,Buonanno:2006ui,Berti:2007fi,Schnittman:2007ij,Baker:2008mj,Pan:2011gk,Kelly:2011bp,Kelly:2012nd}. 
However, in the extreme and small mass-ratio regime, other QNMs can be excited~\cite{Krivan:1997hc,Dorband:2006gg,Nagar:2006xv,Damour:2007xr,Bernuzzi:2010ty,Barausse:2011kb}. We have found that, for $\ell = m$ modes, the QNM mixing is present when $q\leq0$ (see Fig.~\ref{fig:a_-07_22}), and arises mainly due to modes with opposite $m$, whose excitation grows as the spin decreases; for negative spins, the orbit changes direction during plunge (since the particle eventually locks to the rotating BH horizon), thus exciting $(\ell, -m,0)$ modes. For $\ell \neq m$ modes, instead, we have found QNM mixing across the entire spin range. For the $(2,1)$ mode, the main source of mixing is the $(2,-1,0)$ QNM. For the $(3,2)$ mode (see Fig.~\ref{fig:Stack32}), we have recognized 3 different behaviors: when $q\gtrsim0.8$ the ringdown is dominated by $(2,2,0)$ with contamination from $(3,2,0)$; when $0<q\lesssim0.7$, the ringdown is dominated by $(3,2,0)$ with contamination from $(2,2,0)$; when $q\leq0$, the ringdown is dominated by $(3,2,0)$ with contamination from both $(3,-2,0)$ and $(2,2,0)$. The excitation of QNMs with the same $m$, but with different $\ell$, is understood as a basis effect, since the QNMs are computed in a $-2$-spin-weighted spheroidal-harmonic separation of the Teukolsky equation, while the waveforms used in modeling are decomposed in $-2$-spin-weighted spherical-harmonic modes. We have fitted the relative amplitude between the main QNMs that are interfering for each mode (see Tables~\ref{tab:Oppositem} and \ref{tab:Oddmodes}), and have been 
able to model the ringdown Teukolsky waveforms using Eq.~(\ref{BothRD}) for all spins except $q=-0.95, -0.99$ 
for all modes, and also $q=-0.9$ for the (3,2) mode.

In Sec.~\ref{sec:EOBmodeling}, we have discussed how the inspiral-merger-ringdown Teukolsky waveforms helped the extension of the comparable-mass EOB model for nonprecessing, spinning BH binaries of Ref.~\cite{Taracchini:2013rva} to small mass ratios and large spins. In particular, a time delay $\Delta t^{22}_{\rm peak}$ was introduced between the orbital frequency peak $t_{\rm peak}^{\Omega}$ and the point $t_{\rm peak}^{\Omega}+\Delta t_{\rm peak}^{22}$ where non-quasicircular corrections are applied to the merger waveform. The specific dependence of the time-delay function on the spin takes into account the extreme circularity of the orbits encountered in the test-particle limit for large spins, and guarantees that $t_{\rm peak}^{\Omega}+\Delta t_{\rm peak}^{22}$ always lies in a region with significant radial motion. Older EOB models took $t_{\rm peak}^{\Omega}+\Delta t_{\rm peak}^{22}$ to coincide with the peak of the amplitude; however, in this paper, we have shown that such prescription is not adequate in the test-particle limit and, more generally when the mass ratio is smaller than $\sim 1/100$ if $q>0.8$, since the peak occurs much before the ISCO and light-ring. The Teukolsky waveforms were also exploited to build fitting functions for the input values (i.e., $|h_{22}|$, $\partial_{t}^{2}|h_{22}|$, $\omega_{22}$, $\partial_{t}\omega_{22}$ at a point in time during merger) which are needed to impose non-quasicircular corrections to the merger EOB waveform. 

Finally, in Sec.~\ref{sec:CompareWaveforms}, we have evaluated the comparable-mass EOB model of Ref.~\cite{Taracchini:2013rva} in the test-particle limit, and compared it to Teukolsky waveforms computed along the same EOB trajectory. We have found that, up to a spin $\sim 0.5$, the EOB waveforms (based on the factorized resummation of PN formulae in Refs.~\cite{Damour:2008gu,Pan:2010hz}) perform well, with phase differences within 0.1 rads and amplitude errors which are negligible up to merger (see Fig.~\ref{fig:SEOBNRv2_05}). For larger spins, instead, while the EOB model can produce a reasonable $(2,2)$ mode waveform (see Fig.~\ref{fig:SEOBNRv2_08}), still it disagrees with the Teukolsky data, due to the poor performance of the current factorized waveforms in such highly relativistic regimes --- for example for $q=0.8$, we find an amplitude error around 5\% during the late inspiral and a 
dephasing of about 0.8 rads 50 GW cycles before merger, and growing as one moves to lower frequencies. We have also discussed the limitations of the current factorized EOB energy flux, namely the lack of horizon-absorption terms and the truncation at $\ell=8$ modes.

The natural extension of this project will consider inclined orbits in Kerr spacetime. Even at the level of geodetic motion, there exist orbits with constant separation and inclination (with respect to the direction of the Kerr spin), which display precession of the orbital plane. Thus, these orbits will radiate waveforms which carry amplitude
and phase modulations due to the precession. On the analytical side,
we have shown in this paper several limitations of the current EOB
  factorized flux~\cite{Damour:2008gu,Pan:2010hz} for large spins. Thus, it will be crucial to
  improve this flux in the future either by designing a new
  resummation scheme, or by incorporating higher-order PN terms that
  have been recently computed~\cite{Shah:2014tka}. Moreover, the current EOB flux was
developed for nonprecessing BH binaries only; we plan to test
different prescriptions that could extend its validity to the
precessing case. Such work can help the more challenging EOB modeling
of precessing, comparable-mass BH binaries, which has first been
tackled in Ref.~\cite{Pan:2013rra}.

\begin{acknowledgments}
We thank Enrico Barausse, Yi Pan, Achilleas Porfyriadis, and Nico Yunes for useful, informative discussions. 

A.B. and A.T. acknowledge partial support from NSF Grants
No. PHY-0903631 and No. PHY-1208881.  A.B. also acknowledges partial
support from the NASA Grant NNX12AN10G and A.T. from the Maryland
Center for Fundamental Physics. This work was supported at MIT by NSF Grant PHY-1068720.  
G.K. acknowledges research support from NSF Grants No. PHY-1016906, No. CNS-0959382, 
No. PHY-1135664, and No. PHY-1303724,  and from the U.S. Air Force Grant No. FA9550-10-1-0354 
and No. 10-RI-CRADA-09.
\end{acknowledgments}

\appendix
\section{Input values for non-quasicircular corrections to merger waveforms}
\label{app:InputValues}
\begin{figure}[!ht]
  \begin{center}
    \includegraphics[width=0.5\textwidth]{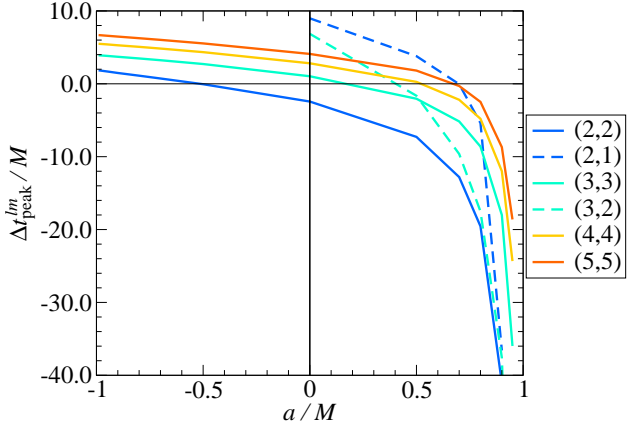}
\caption{\label{fig:Deltatpeak} Time delay between the orbital frequency peak and the Teukolsky amplitude peak, defined as $\Delta t_{\rm peak}^{\ell m} \equiv t_{\rm peak}^{\ell m} - t_{\rm peak}^{\Omega}$. The value of $\Delta t_{\rm peak}^{22}$ for spin 0.95 is $-103M$, and exceeds the range of the plot.}
  \end{center}
\end{figure}
\begin{figure}[!ht]
\begin{center}
\includegraphics[width=0.5\textwidth]{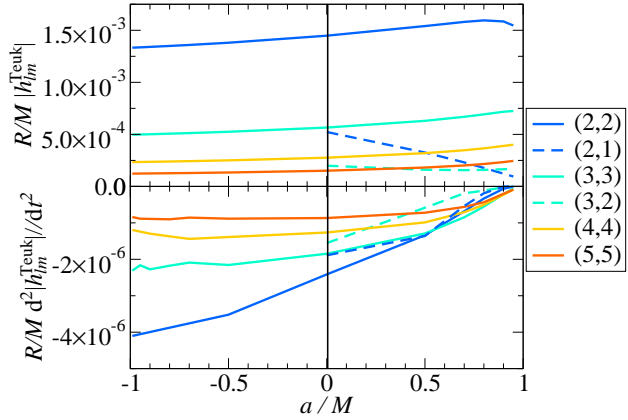} 
\caption{\label{fig:IVAmp} Amplitude and curvature of the Teukolsky waveforms at their amplitude peak. $R$ is the distance to the source.}
  \end{center}
\end{figure}
\begin{figure}[!ht]
%\vskip1cm
\begin{center}
\includegraphics[width=0.5\textwidth]{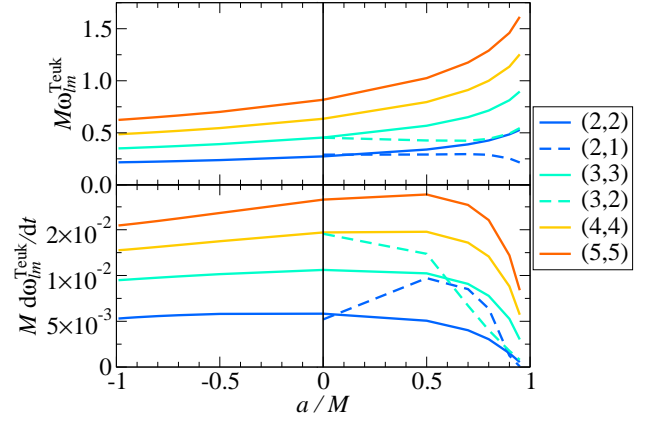}
\caption{\label{fig:IVFreq} Frequency and derivative of the frequency of the Teukolsky waveforms at their amplitude peak.}
  \end{center}
\end{figure}
In this appendix, we provide useful information about the Teukolsky merger waveforms that can be exploited in the construction of comparable-mass, spinning, nonprecessing EOB models that span the entire physical parameter space, as discussed in Sec.~\ref{sec:EOBmodeling}. We omit spin 0.99 because it is difficult to determine its peak positions $t_{\rm peak}^{\ell m}$, due to the extreme flatness of the mode amplitudes, as shown in Sec.~\ref{sec:Flat}. We also omit the negative spins for the $(2,1)$ and $(3,2)$ modes since, as discussed in Sec.~\ref{sec:odd}, QNM mixing has an early onset (around the turning point of the azimuthal motion for $(2,1)$; slightly later than that for $(3,2)$), and affects the peak of the waveform; it is therefore ambiguous where to measure the input values for these cases.

In Fig.~\ref{fig:Deltatpeak} we show how the time delay between the orbital frequency peak $t_{\rm peak}^{\Omega}$ and the Teukolsky amplitude peak $t_{\rm peak}^{\ell m}$ changes with the Kerr spin. As pointed out in Sec.~\ref{sec:Flat}, the amplitudes tend to peak earlier and earlier as $q$ increases, well before the ISCO when $q>0.8$. This creates difficulties when applying the non-quasicircular procedure to correct the EOB merger waveforms at $t_{\rm peak}^{\ell m}$, as elucidated in Sec.~\ref{sec:EOBmodeling}. In fact, in the comparable-mass EOB model of Ref.~\cite{Taracchini:2013rva}, we chose a delay $\Delta t_{\rm peak}^{22}$ which decreases after spin 0.8, thus departing from the blue curve in Fig.~\ref{fig:Deltatpeak}. 

In Figs.~\ref{fig:IVAmp} and \ref{fig:IVFreq} we plot the input values computed at the time $t^{\ell m}_{\rm peak}$ when the Teukolsky amplitudes peak. The largest numerical uncertainties are visible on the curvature, but, as it turns out, the EOB waveforms are only mildly sensitive to such input value; in order to get a good modeling, the crucial input values are rather the values of the amplitude and the frequency.

\bibliography{References/References}

\end{document}